\documentclass{article}
\usepackage{fancyhdr}
\usepackage{report}

\usepackage{soul}
\usepackage[utf8]{inputenc} % allow utf-8 input
\usepackage[T1]{fontenc}    % use 8-bit T1 fonts
\usepackage{hyperref}       % hyperlinks
\usepackage{url}            % simple URL typesetting
\usepackage{booktabs}       % professional-quality tables
\usepackage{amsfonts}       % blackboard math symbols
\usepackage{fancyhdr}       % custom headers and footers

\usepackage{nicefrac}       % compact symbols for 1/2, etc.
\usepackage{microtype}      % microtypography
\usepackage{graphicx}
\usepackage{pifont}
\usepackage{multirow}
\usepackage{makecell}
\usepackage{CJKutf8}
\definecolor{darkmagenta}{rgb}{0.56, 0.0, 1.0}
\definecolor{softyellow}{rgb}{1.0, 0.92, 0.3} % richer, warmer yellow
\definecolor{LightAquamarine}{rgb}{0.75, 1.0, 0.8} % soft aqua green
\definecolor{FireBrick}{RGB}{178,34,34}
\definecolor{MediumPurple}{RGB}{147,112,219}

\definecolor{uclablue}{rgb}{0.15, 0.45, 0.68}
\hypersetup{
    breaklinks,
    colorlinks=true,
    citecolor={darkmagenta},
    linkcolor={uclablue},
    urlcolor={uclablue}
}
\usepackage{float}
\usepackage{subcaption}
\usepackage{placeins}
\usepackage{tcolorbox}
\usepackage{amsmath}
\usepackage{amssymb}
\usepackage{xspace}
\usepackage{enumitem}
\usepackage{multirow} 
\tcbuselibrary{breakable}
\usepackage{enumitem}
\usepackage{colortbl}
\usepackage{fancyhdr}

\usepackage{tcolorbox}
\usepackage{pgfplots}
\usepgfplotslibrary{polar}
\pgfplotsset{compat=1.18}

\tcbuselibrary{most}
\definecolor{highlight}{RGB}{230, 210, 255}
\usepackage{tcolorbox}
\usepackage{pgfplots}
\usepgfplotslibrary{polar}
\pgfplotsset{compat=1.18}

\tcbuselibrary{most,listings}

\definecolor{promptbg}{RGB}{248,249,252}
\definecolor{promptframe}{RGB}{52,110,183}
\definecolor{promptsep}{RGB}{180,200,230}

\newtcblisting{promptbox}[1]{%
  listing only,
  colback=promptbg,
  colframe=promptframe,
  title={\small\bfseries #1},
  fonttitle=\small\bfseries,
  listing options={%
    basicstyle=\ttfamily\scriptsize,
    breaklines=true,
    breakautoindent=false,
    columns=fullflexible,
    keepspaces=true,
    aboveskip=0pt,
    belowskip=0pt,
  },
  boxrule=0.6pt,
  arc=2pt,
  left=6pt, right=6pt, top=2pt, bottom=2pt,
  breakable,
}

\definecolor{highlight}{RGB}{230, 210, 255}

\newtcbox{\tagtext}{
  on line, arc=3pt, colback=yellow!30, colframe=yellow!90,
  boxsep=0pt, left=4pt, right=4pt, top=2pt, bottom=2pt,
  boxrule=0.8pt, fontupper=\scriptsize\bfseries, nobeforeafter
}
\newtcbox{\tagimage}{
  on line, arc=3pt, colback=green!10, colframe=green!60!black,
  boxsep=0pt, left=4pt, right=4pt, top=2pt, bottom=2pt,
  boxrule=0.8pt, fontupper=\scriptsize\bfseries\color{green!50!black}, nobeforeafter
}
\newtcbox{\tagvideo}{
  on line, arc=3pt, colback=red!8, colframe=red!60,
  boxsep=0pt, left=4pt, right=4pt, top=2pt, bottom=2pt,
  boxrule=0.8pt, fontupper=\scriptsize\bfseries\color{red!70!black}, nobeforeafter
}
\definecolor{njuPurple}{RGB}{220,205,230}     % 南大紫（深）
\definecolor{njuPurpleLight}{RGB}{250,245,252}   % 极浅的紫色背景（接近白）

\newtcolorbox{abstractbox}{
    colback=njuPurpleLight,   % 浅紫色背景
    colframe=njuPurple,       % 深紫色边框
    boxrule=1pt,              % 边框粗细
    arc=4mm,                  % 圆角
    left=8pt,                 % 左边距
    right=8pt,                % 右边距
    top=8pt,                  % 上边距
    bottom=8pt,               % 下边距
    opacityback=0.95
}
\usepackage{caption}
\usepackage{cuted}

\definecolor{headerblue}{RGB}{40, 80, 140}
\definecolor{headerbg}{RGB}{220, 235, 255}
\definecolor{closedtag}{RGB}{245, 240, 255}
\definecolor{opentag}{RGB}{240, 255, 240}
\definecolor{rowgray}{RGB}{248, 248, 252}
\definecolor{highlight}{RGB}{220, 190, 255}
\definecolor{best}{RGB}{0, 140, 100}
\definecolor{second}{RGB}{30, 100, 180}

\definecolor{genhead}{RGB}{52, 110, 183}     % steel blue
\definecolor{genbg}{RGB}{227, 238, 252}       % light blue
\definecolor{edithead}{RGB}{46, 139, 87}      % sea green
\definecolor{editbg}{RGB}{228, 248, 235}      % light green
\definecolor{repairhead}{RGB}{192, 72, 54}    % terracotta
\definecolor{repairbg}{RGB}{252, 232, 228}    % light red
\definecolor{rowalt}{RGB}{248, 249, 252}      % alternating row background

\newcommand{\best}[1]{\textcolor{best}{\textbf{#1}}}
\newcommand{\second}[1]{\textcolor{second}{\underline{#1}}}

\title{WebCompass: Towards Multimodal Web Coding Evaluation  for Code Language Models}

\author{
Xinping Lei$^{\dagger}$, Xinyu Che$^{\dagger}$, Junqi Xiong$^{\dagger}$, Chenchen Zhang$^{\dagger}$, Yukai Huang$^{\dagger}$, Chenyu Zhou$^{\dagger}$,
Haoyang Huang, Minghao Liu, Letian Zhu, Hongyi Ye, Jinhua Hao, Ken Deng, Zizheng Zhan, Han Li, Dailin Li,
Yifan Yao, Ming Sun, Zhaoxiang Zhang, Jiaheng Liu$^{*}$\\[2pt]
\small Nanjing University \quad Kuaishou Technology\\
\small $^{\dagger}$Equal contribution.\quad $^{*}$Corresponding author.
}
\date{}

\begin{document}

\maketitle

\begin{abstract}

Large language models are rapidly evolving into interactive coding agents capable of end-to-end web coding, yet existing benchmarks evaluate only narrow slices of this capability---typically text-conditioned generation with static-correctness metrics---leaving visual fidelity, interaction quality, and codebase-level reasoning largely unmeasured.
We introduce \textbf{WebCompass}, a comprehensive, multimodal benchmark that provides a \emph{unified lifecycle evaluation} of web engineering capability.
Recognizing that real-world web coding is an iterative cycle of generation, editing, and repair, WebCompass spans three input modalities (\textbf{text, image, and video}) and three tightly coupled task types (\textbf{generation, editing, and repair}), yielding seven complementary task categories that closely mirror professional workflows.
Through a multi-stage, human-in-the-loop pipeline, we curate high-quality instances covering 15 generation domains, 16 editing operation types, and 11 repair defect types, each annotated at Easy/Medium/Hard difficulty levels.
On the evaluation side, we adopt a \emph{checklist-guided LLM-as-a-Judge} protocol for editing and repair, and propose a novel \textbf{Agent-as-a-Judge} paradigm for generation that autonomously executes generated websites in a real browser, explores interactive behaviors via the Model Context Protocol (MCP), and iteratively synthesizes targeted test cases---closely approximating human acceptance testing.
We evaluate a diverse set of representative closed-source and open-source models and observe that:
(1)~closed-source models remain substantially stronger and more balanced;
(2)~editing and repair exhibit distinct difficulty profiles, with repair preserving interactivity better but remaining execution-challenging;
(3)~aesthetics is the most persistent bottleneck, especially for open-source models; and
(4)~framework choice materially affects outcomes, with Vue consistently challenging while React and Vanilla/HTML perform more strongly depending on task type.
All benchmark data\footnote{\url{https://huggingface.co/datasets/NJU-LINK/WebCompass}}, evaluation code\footnote{\url{https://github.com/NJU-LINK/WebCompass}}, and project page\footnote{\url{https://nju-link.github.io/WebCompass/}} are publicly available.

\end{abstract}

\section{Introduction}

\begin{figure*}[h]
    \centering
    \begin{minipage}[b]{0.29\textwidth}
        \centering
        \includegraphics[width=\linewidth]{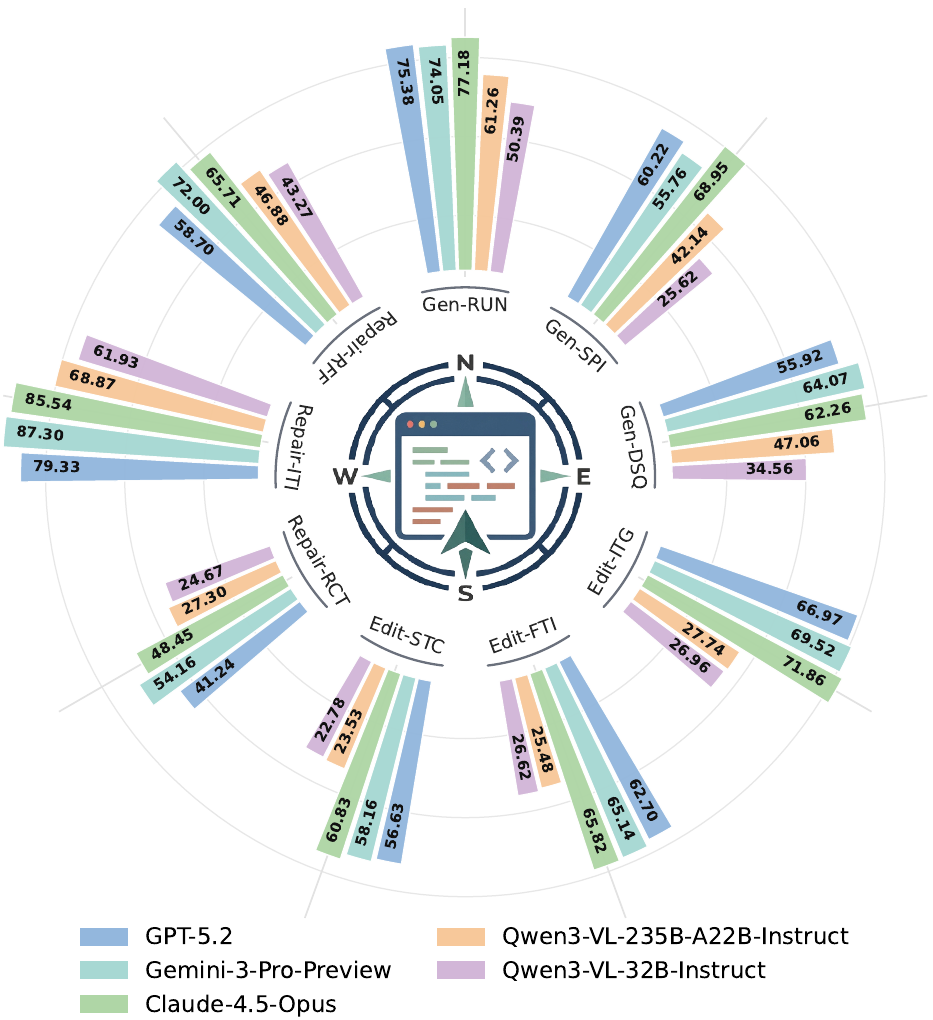}
        \caption{Radar chart of model performance across all seven task types in WebCompass.}
        \label{fig:compass}
    \end{minipage}
    \hfill
    \begin{minipage}[b]{0.70\textwidth}
        \centering
        \includegraphics[width=\linewidth]{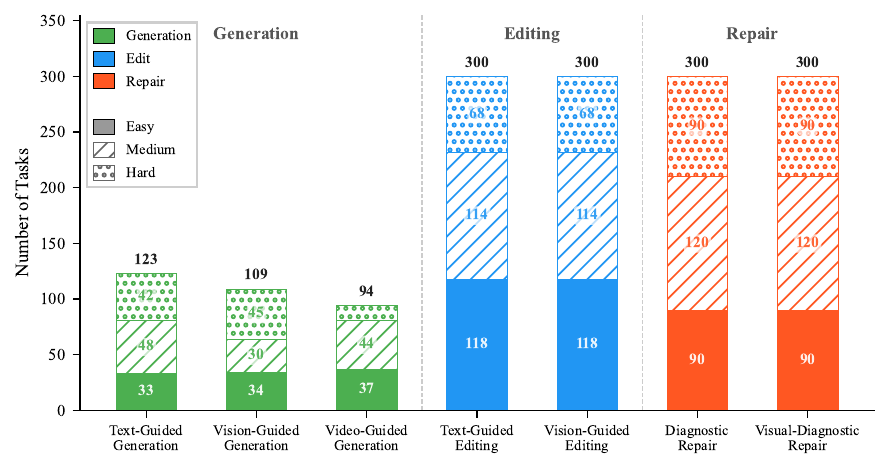}
        \caption{Difficulty distribution of WebCompass. }
        \label{fig:difficulty_distribution}
    \end{minipage}
\end{figure*}

% ---- Comparison Table ----
\begin{table*}[htbp]
\centering
\caption{Comparison with prior web coding benchmarks. WebCompass is the first to support all three task types across text, image, and video modalities. \textbf{Gen.}=Generation, \textbf{Edit}=number of supported editing categories, \textbf{Rep.}=number of supported repair categories, \textbf{Multi-page}=project-level multi-page testing, \textbf{Interact.}=interactive functionality evaluation, \textbf{Visual}=aesthetics and visual fidelity evaluation, \textbf{Agentic}=Agent-as-a-Judge dynamic testing (using LLM agents to interact with browsers and synthesize tests), \textbf{Reverse}=reverse-engineered deterministic repair tasks. A red cross indicates that the task family is not supported. Data sizes are reported as the number of tasks or question-answer pairs.}
\label{tab:benchmark_comparison}
\resizebox{\textwidth}{!}{%
\renewcommand{\arraystretch}{1.2}
\begin{tabular}{@{}llccccccccl@{}}
\toprule
\textbf{Benchmark} & \textbf{Size} & \textbf{Gen.} & \textbf{Edit (\#)} & \textbf{Rep. (\#)} & \textbf{Multi-page} & \textbf{Interact.} & \textbf{Visual} & \textbf{Agentic} & \textbf{Reverse} & \textbf{Input Modality} \\
\midrule
\multicolumn{11}{c}{\cellcolor{closedtag}\textit{\textcolor{headerblue}{\textbf{Generation-Only Benchmarks}}}} \\
\midrule
Interaction2Code \citep{wan2024interaction2code} & 504 & \textcolor{teal}{\ding{51}} & \textcolor{red!70}{\ding{55}}& \textcolor{red!70}{\ding{55}}& \textcolor{red!70}{\ding{55}} & \textcolor{teal}{\ding{51}} & \textcolor{teal}{\ding{51}} & \textcolor{red!70}{\ding{55}} & \textcolor{red!70}{\ding{55}} & \tagimage{Image} \\
\rowcolor{rowgray}
FronTalk \citep{wu2025frontalk} & 1000 & \textcolor{teal}{\ding{51}} & \textcolor{red!70}{\ding{55}}& \textcolor{red!70}{\ding{55}}& \textcolor{teal}{\ding{51}}& \textcolor{teal}{\ding{51}}& \textcolor{teal}{\ding{51}}& \textcolor{red!70}{\ding{55}} & \textcolor{red!70}{\ding{55}} & \tagtext{Text}\;\tagimage{Image} \\
Web-Bench \citep{xu2025web} & 1000 & \textcolor{teal}{\ding{51}} & \textcolor{red!70}{\ding{55}}& \textcolor{red!70}{\ding{55}}& \textcolor{teal}{\ding{51}} & \textcolor{teal}{\ding{51}} & \textcolor{red!70}{\ding{55}} & \textcolor{red!70}{\ding{55}} & \textcolor{red!70}{\ding{55}} & \tagtext{Text}\;\tagimage{Image} \\
\rowcolor{rowgray}
FrontendBench \citep{zhu2025frontendbench} & 148 & \textcolor{teal}{\ding{51}} & \textcolor{red!70}{\ding{55}}& \textcolor{red!70}{\ding{55}}& \textcolor{red!70}{\ding{55}} & \textcolor{teal}{\ding{51}}& \textcolor{red!70}{\ding{55}}& \textcolor{red!70}{\ding{55}} & \textcolor{red!70}{\ding{55}} & \tagtext{Text} \\
WebApp1K \citep{cui2024webapp1k} & 1000 & \textcolor{teal}{\ding{51}} & \textcolor{red!70}{\ding{55}}& \textcolor{red!70}{\ding{55}}& \textcolor{teal}{\ding{51}}& \textcolor{red!70}{\ding{55}} & \textcolor{teal}{\ding{51}}& \textcolor{red!70}{\ding{55}} & \textcolor{red!70}{\ding{55}} & \tagtext{Text}\\
\rowcolor{rowgray}
IWR-Bench~\citep{chen2025iwr} & 113 & \textcolor{teal}{\ding{51}} & \textcolor{red!70}{\ding{55}}& \textcolor{red!70}{\ding{55}}& \textcolor{teal}{\ding{51}}& \textcolor{teal}{\ding{51}}& \textcolor{teal}{\ding{51}}& \textcolor{red!70}{\ding{55}} & \textcolor{red!70}{\ding{55}} & \tagvideo{Video} \\
WebGen-Bench \citep{lu2025webgen} & 101 & \textcolor{teal}{\ding{51}} & \textcolor{red!70}{\ding{55}}& \textcolor{red!70}{\ding{55}}& \textcolor{teal}{\ding{51}}& \textcolor{teal}{\ding{51}}& \textcolor{red!70}{\ding{55}}& \textcolor{red!70}{\ding{55}} & \textcolor{red!70}{\ding{55}} & \tagtext{Text} \\
\midrule
\multicolumn{11}{c}{\cellcolor{opentag}\textit{\textcolor{headerblue}{\textbf{Multi-Task Benchmarks}}}} \\
\midrule
SWE-bench MM \citep{yang2024swebenchmultimodal} & 517 & \textcolor{red!70}{\ding{55}} & 3& 4& \textcolor{teal}{\ding{51}} & \textcolor{red!70}{\ding{55}}& \textcolor{red!70}{\ding{55}}& \textcolor{red!70}{\ding{55}} & \textcolor{red!70}{\ding{55}} & \tagtext{Text}\;\tagimage{Image} \\

DesignBench \citep{xiao2025designbench} & 900 & \textcolor{teal}{\ding{51}} & 6& 6& \textcolor{teal}{\ding{51}}& \textcolor{red!70}{\ding{55}} & \textcolor{teal}{\ding{51}} & \textcolor{red!70}{\ding{55}} & \textcolor{red!70}{\ding{55}} & \tagimage{Image} \\
\midrule
\rowcolor{highlight}
\textbf{WebCompass (Ours)} & \textbf{1526} & \textcolor{teal}{\ding{51}} & \textbf{16} & \textbf{11} & \textcolor{teal}{\ding{51}} & \textcolor{teal}{\ding{51}} & \textcolor{teal}{\ding{51}} & \textcolor{teal}{\ding{51}} & \textcolor{teal}{\ding{51}} & \tagtext{Text}\;\tagimage{Image}\;\tagvideo{Video} \\
\bottomrule
\end{tabular}%
}
\end{table*}

Large Language Models (LLMs) have rapidly evolved from passive code assistants into interactive coding agents capable of implementing substantial software changes from natural-language instructions~\citep{yang2024sweagent, wang2024openhands, cognition2024devin}. This progress is especially evident in \textbf{web development}, where outputs can be directly executed, visually inspected, and iteratively refined. A growing body of work has proposed benchmarks that span different task types and input modalities for web coding (Table~\ref{tab:benchmark_comparison}).

Yet evaluating web coding is fundamentally different from evaluating traditional code generation. Success depends not only on functional correctness, but also on visual fidelity, interaction behavior, responsiveness, accessibility, and overall user experience. These aspects are difficult to capture with standard code-centric metrics such as pass@k on HumanEval~\citep{chen2021evaluating} or unit-test pass rates on SWE-Bench~\citep{jimenez2023swe}, which focus on algorithmic correctness or repository-level bug fixing rather than interactive front-end applications.

To address this gap, we introduce \textbf{WebCompass}, a unified multimodal benchmark and evaluation framework for web coding. WebCompass spans \textbf{text, image, and video} inputs, covers \textbf{generation, editing, and repair} tasks, and adopts \textbf{task-aware evaluation} tailored to each setting. For editing and repair, we use a \emph{checklist-guided LLM-as-a-Judge} protocol~\citep{zheng2023judging}, which is well suited to patch-based outputs with constrained solution spaces. For generation, we propose an \textbf{Agent-as-a-Judge} protocol~\citep{zhuge2024agent}, in which an autonomous agent launches the generated website in a real browser, explores it through MCP, synthesizes targeted test cases, and scores the result based on execution.

This design reflects the differing nature of web coding tasks. Editing and repair are localized and checklist-aligned, making diff-level inspection and before/after screenshots sufficient for reliable evaluation. Generation, by contrast, is open-ended and long-horizon, with correctness often depending on multi-step runtime behavior that static inspection cannot capture. By combining multimodal task coverage with execution-based evaluation, WebCompass provides a more realistic and scalable benchmark for assessing web coding agents.

\paragraph{Contributions.}
\textbf{(1) Unified lifecycle coverage.} Unlike prior benchmarks that target isolated tasks or modalities (Table~\ref{tab:benchmark_comparison}), WebCompass jointly evaluates generation, editing, and repair across text, image, and video inputs, enabling cross-task and cross-modality comparisons within a single framework.
\textbf{(2) Rigorous and deterministic task construction.} We refine underspecified queries into structured design documents for generation, synthesize context-consistent requirements without leaking implementation details for editing, and provide exact \texttt{search/replace} annotations mapping buggy code to clean targets for repair, ensuring reproducible evaluation.
\textbf{(3) Task-aware evaluation paradigms.} We introduce an Agent-as-a-Judge protocol that combines real-browser interaction with iterative test-case synthesis for open-ended generation tasks, complementing checklist-guided LLM-as-a-Judge for constrained patch-based tasks.

\section{WebCompass}
\label{sec:benchmark_data}

\subsection{Overview}
\begin{figure*}[t]
    \centering
    \includegraphics[width=1\linewidth]{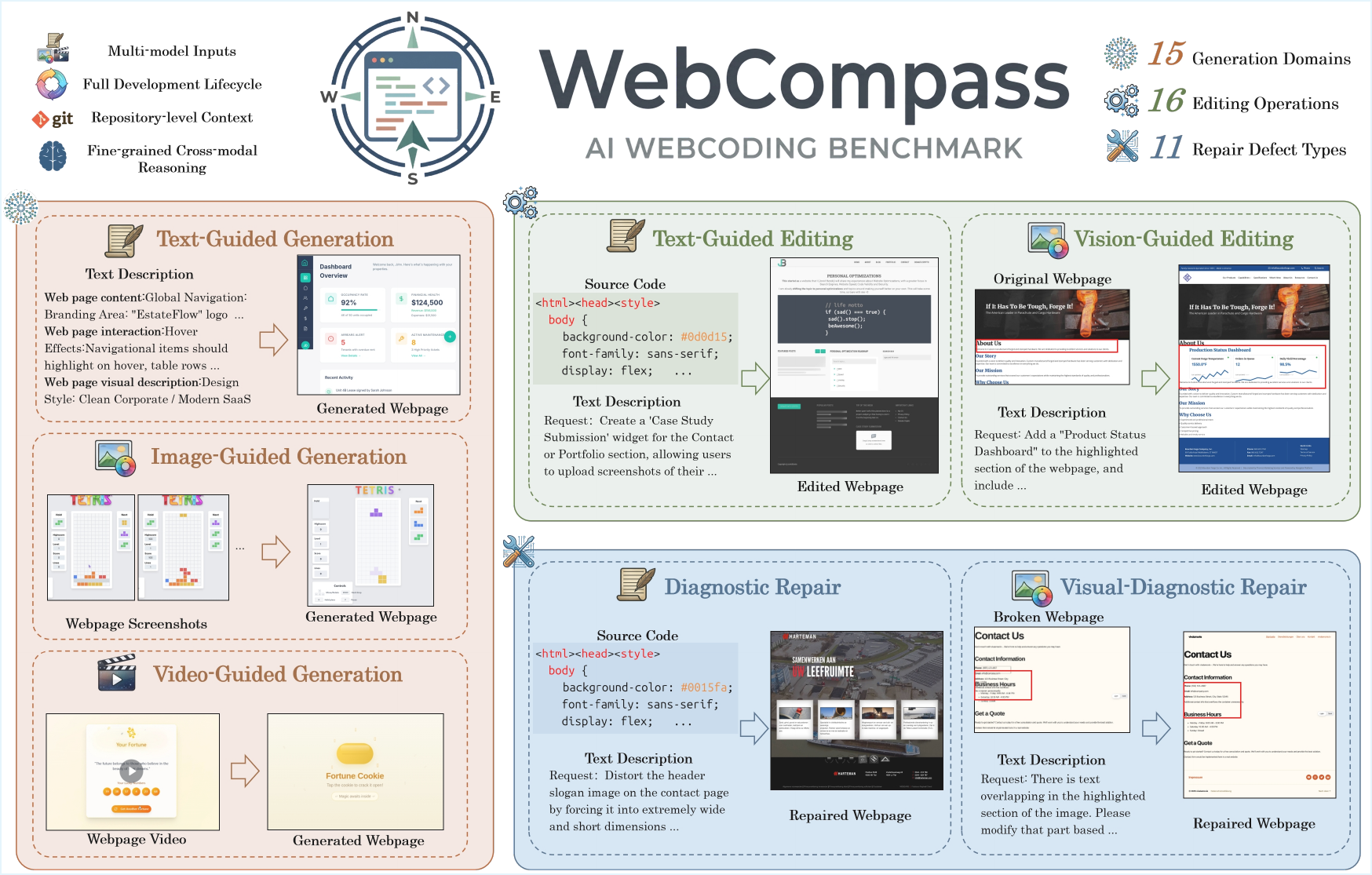}
    \caption{Overview of WebCompass. The benchmark supports three input modalities (text, image, video) and three task types (generation, editing, repair), resulting in seven complementary task categories that cover the full lifecycle of web development.}
    \label{fig:overview}
\end{figure*}

WebCompass supports three input modalities (text, image, and video) and three types of web coding tasks (generation, editing, and repair), resulting in seven task categories: \textit{Text-Guided Generation} (text-conditioned web generation), \textit{Vision-Guided Generation} (image-conditioned web generation), \textit{Video-Guided Generation} (video-conditioned web generation), \textit{Text-Guided Editing} (text-instructed web editing via patches), \textit{Vision-Guided Editing} (image-grounded web editing via patches), \textit{Diagnostic Repair} (text-described web repair via patches), and \textit{Visual-Diagnostic Repair} (image-grounded web repair via patches). Each task is designed to closely reflect real-world development scenarios. We define each task as follows:

\begin{enumerate}
    \item \textbf{Text-Guided Generation.} The input is a textual specification of a target web page, consisting of three aspects: (i) page content, (ii) interaction behaviors, and (iii) visual appearance. The model is required to output a complete web code repository that satisfies the specification.

    \item \textbf{Vision-Guided Generation.} The input comprises multiple screenshots of a web page. Beyond presenting content, layout, and visual styling, the screenshots are also intended to capture interactive functionalities. Depending on the data source, we consider two types of screenshot sets: (i) a collection covering the main page and its subpages, and (ii) a sequence capturing page state changes during browsing. The model is required to reproduce a web code repository whose visual appearance and functionality match the screenshots.

    \item \textbf{Video-Guided Generation.} The input is a screen-recorded browsing video containing multiple user interactions. The model is required to generate a web code repository whose appearance and functionality are consistent with those demonstrated in the video.

    \item \textbf{Text-Guided Editing.} The input includes a web code repository and a text-based editing instruction. The model is required to output a \emph{code patch} that edits the repository such that the updated web page meets the instructions.

    \item \textbf{Vision-Guided Editing.} The input includes a screenshot of the current web page, the corresponding web code repository, and an editing instruction. The model is required to output a \emph{code patch} that modifies the repository so that the edited web page satisfies the instruction.

    \item \textbf{Diagnostic Repair.} The input includes a web code repository and a textual description of the existing issues. The model is required to output a \emph{code patch} that repairs the repository and resolves the described problems.

    \item \textbf{Visual-Diagnostic Repair.} The input includes a screenshot of the current web page, the web code repository, and a description of the existing issues. The model is required to output a \emph{code patch} that repairs the repository and resolves the described problems.
\end{enumerate}

Taken together, WebCompass serves as a comprehensive benchmark to evaluate the capabilities of multimodal models in realistic web engineering scenarios.
Beyond basic code generation, it rigorously assesses a model's proficiency across several critical dimensions:
(1) \textbf{Nuanced User Intent Understanding}, encompassing layout structure, aesthetic design styles, and complex interaction logic;
(2) \textbf{Fine-grained Cross-modal Reasoning}, requiring precise alignment between visual inputs (images/videos) and code implementations;
(3) \textbf{Repository-level Context Awareness}, testing the ability to maintain consistency within existing codebases during editing and repairing;
and (4) \textbf{Diagnostic \& Problem-Solving Skills}, specifically for identifying and fixing semantic or visual anomalies.

\subsection{Data Collection}
\label{subsec:data_collection}
To ensure the benchmark reflects real-world scenarios, we employ a multi-stage, human-in-the-loop pipeline to construct a high-quality benchmark covering all seven task types. Figure~\ref{fig:data_construction_pipeline} illustrates the overall process.

\subsubsection{Text-Guided Generation.} We design the Text-Guided Generation set to (i) contain realistic and actionable requirements and (ii) cover diverse web page types. We therefore collect initial queries from multiple complementary sources: WebGen-Bench~\citep{lu2025webgen} (manually constructed queries), ArtifactsBench~\citep{zhang2025artifactsbench} (diverse page categories with rigorous filtering), BigCode Arena (real user requests), and high-quality web showcases from V0 (an AI IDE for web coding). These sources form our initial query pool. To reduce redundancy, we embed queries using BGE-M3 and perform $k$-means clustering to obtain a deduplicated candidate set. We then use an LLM to assign category and difficulty labels to each query (five independent annotations per query), taking the majority vote as the final label. Finally, we perform stratified sampling across categories and difficulties to obtain 123 text-guided generation queries.

However, we observe that queries from everyday usage scenarios are often \emph{underspecified}, leading to large variations in generated pages across models. While such low-constraint queries can test a model's creativity, they hinder automated evaluation because creativity and implicit-intent matching are subjective and difficult to judge automatically---it is unclear whether the model is being ``overly clever'' or truly aligned with user intent. To address this, we prompt an LLM to act as a product manager and elaborate each underspecified request into a structured web design document covering (1) page content, (2) interaction behaviors, and (3) visual appearance.

\subsubsection{Vision-Guided Generation.}
Although many existing datasets include webpage screenshots, most contain relatively simple UIs that are insufficient to challenge modern models. We observe that WebRenderBench provides a large number of visually complex webpages, but typically only includes a single screenshot per website. We thus perform data augmentation: we parse the subpage URLs referenced in \texttt{index.html}, randomly select two, and use Playwright to capture their screenshots. To further test whether models can reproduce multi-page websites and their dependency relationships, we inject a JavaScript overlay into the main-page screenshot to highlight the positions of subpage URLs with colored bounding boxes. Due to network instability and dynamic content loading, screenshots may contain artifacts. We therefore conduct multiple rounds of LLM-based verification as an initial filter, followed by manual inspection.

In addition, most existing datasets only provide \emph{static} screenshots and lack dynamic webpage content. Although Interaction2Code~\citep{wan2024interaction2code} supplies multiple images to convey certain interaction information, it still cannot adequately represent animations and complex interaction patterns. To fill this gap, we browse diverse webpages from V0 and Figma and manually extract keyframes capturing critical state changes. These two components---augmented multi-page screenshots and dynamic keyframe sequences---together constitute the Vision-Guided Generation test set.

\subsubsection{Video-Guided Generation.} Compared to text and images, videos can more clearly convey dynamic effects such as animations and multi-step interactions. To emphasize this advantage, we manually select webpages from V0 and Figma with rich dynamic behaviors across different categories, browse them, and record interaction videos. Annotators are instructed to first explore each webpage, plan a comprehensive exploration path, and then conduct the final recording to ensure thorough coverage of all interactive features.

\subsubsection{Editing \& Repair Task Data Collection Pipeline}

\begin{figure*}[t]
    \centering
    \includegraphics[width=1\linewidth]{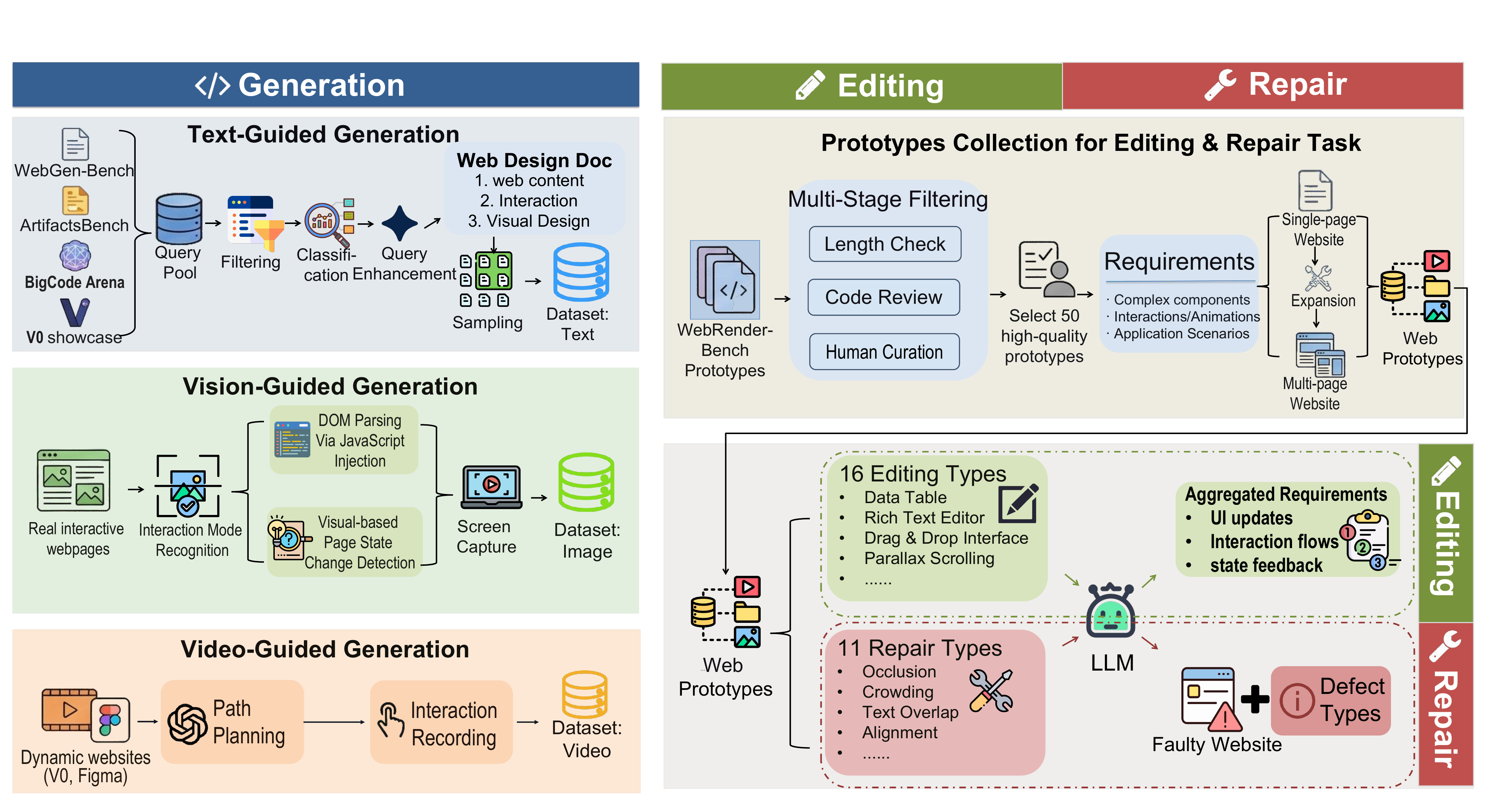}
    \caption{Data construction pipeline for WebCompass.
    \textbf{Top:} prototypes are collected through multi-stage filtering, manual selection, and page-level expansion.
    \textbf{Bottom:} each prototype is converted into editing tasks (left, green) or repair tasks (right, red) following task-type--specific procedures.}
    \label{fig:data_construction_pipeline}
\end{figure*}

\textbf{Prototype Collection for Editing \& Repair.}
Both editing and repair tasks share a common pool of high-quality web prototypes (Figure~\ref{fig:data_construction_pipeline}, top).
We construct these prototypes from the WebRenderBench test set via a three-stage pipeline:
\emph{length filtering} $\rightarrow$ \emph{automatic quality scoring} $\rightarrow$ \emph{human curation}, then expand each selected prototype into single-page and multi-page variants.

\begin{itemize}[leftmargin=1.2em]
    \item \textbf{Stage 1: Length filtering.}
    We constrain the total character count across all code files to 32k--64k, with each individual file no longer than 48k characters.
    These bounds approximate the multi-file coordination complexity of medium-to-large front-end projects, while avoiding overly small instances (lacking difficulty) or overly large ones (inducing context truncation and unstable evaluation).
    \item \textbf{Stage 2: Quality scoring.}
    For candidates satisfying the length constraints, we use GPT-4o to perform a code review on a 10-point scale and retain those scoring $\ge 9$, yielding 81 candidates.
    \item \textbf{Stage 3: Human curation and expansion.}
    We manually select 50 high-quality prototypes.
    Each prototype is kept as a \textsc{Single-page} website and additionally extended into a \textsc{Multi-page} website by adding extra pages, inter-page navigation, and shared resources.
    Together, the two variants constitute the \emph{Web Prototypes} used for all downstream task construction.
\end{itemize}

\textbf{Text-Guided and Vision-Guided Editing.}
Starting from each web prototype as the executable \emph{source} website, we create editing instances by introducing new or enhanced requirements aligned with 16 predefined high-level task types covering complex components (e.g., data tables, rich-text editors, drag-and-drop interfaces), interaction/animation effects (e.g., parallax scrolling), and holistic application scenarios (Figure~\ref{fig:data_construction_pipeline}, bottom-left).
For every task type, we aggregate requirements that specify \emph{what} to change---including UI updates, interaction flows, and state feedback---while deliberately omitting implementation details (e.g., class names, selectors, or CSS values) to ensure fairness and realism.
The resulting requirements, paired with the source website, form the editing instances; \emph{Vision-Guided} variants additionally supply a reference screenshot in lieu of (or alongside) the textual instruction.

\textbf{Diagnostic and Visual-Diagnostic Repair.}
Repair tasks are constructed in a verifiable \emph{reverse} manner (Figure~\ref{fig:data_construction_pipeline}, bottom-right).
We treat a clean web prototype as the \emph{destination} and use an LLM to inject explainable, observable front-end defects drawn from 11 repair types, producing the faulty \emph{source} website.
The model is then required to repair the source back to the destination.
The injected defects span three dimensions:

\begin{itemize}[leftmargin=1.2em]
    \item \textbf{Visual layout:} occlusion, crowding, text overlap, misalignment, insufficient contrast, overflow, and distorted proportions.
    \item \textbf{Semantics \& structure:} incorrect semantic/nesting structures and missing attributes.
    \item \textbf{Interaction usability:} broken interactions and loss of interactivity.
\end{itemize}

We then generate natural-language repair instructions that provide vague hints about potential defect types or underlying issues, rather than a complete description of the problem, ensuring no implementation details are leaked.
To guarantee determinism and support automatic evaluation, each repair instance includes an exact text-level modification annotation (\texttt{search/replace}) that is the strict inverse of the defect-injection edits.
This design ensures (i)~a uniquely correct, runnable solution, (ii)~reproducible transformation from source to destination, and (iii)~automated verification and error localization.
Throughout, we enforce contextual consistency and the ``specify goals, not methods'' principle.

\textbf{Ecological validity of injected defects.}
The 11 defect categories are not arbitrarily chosen.
They are the product of a systematic analysis of over 200 real-world community submissions on V0 and corresponding GitHub Issues, from which we identified the most frequently occurring front-end anti-patterns.
Each category (e.g., Occlusion, Overflow, Loss of Interactivity) corresponds to a high-frequency failure mode observed in practice.
By grounding our synthetic defects in this empirical distribution, we ensure ecological representativeness---models are tested on the kinds of bugs they are most likely to encounter in real-world web development, rather than on artificial corner cases.

\subsection{Quality Control}
\label{subsec:quality_control}

We apply a multi-layered quality assurance process across all task types:

\textbf{Automated checks.} Before human review, every instance passes through a suite of automated validators: (i) all code repositories must compile and render without fatal errors in a headless Chromium environment; (ii) editing and repair patches must apply cleanly to their respective source repositories; and (iii) repair \texttt{search/replace} annotations are verified to be the exact inverse of the defect-injection edits, guaranteeing a unique, deterministic solution.

\textbf{LLM-assisted screening.} We use an LLM to perform multi-round quality checks on generated requirements and screenshots. For Vision-Guided Generation, the LLM verifies that screenshots are complete (no blank regions, missing assets, or broken layouts caused by network issues). For edit and repair tasks, the LLM checks that natural-language instructions are unambiguous, do not leak implementation details, and are consistent with the underlying code changes.

\textbf{Human curation.} All instances undergo a final round of expert human review. Annotators verify (i) the correctness and completeness of task descriptions, (ii) the visual quality of screenshots and videos, (iii) the appropriateness of difficulty labels (Easy/Medium/Hard), and (iv) the alignment between requirements and ground-truth patches. Instances that fail any criterion are revised or discarded.

\subsection{Dataset Statistics}
\label{subsec:dataset_statistics}

We propose a fine-grained taxonomy for the generation, editing, and repair tasks, as detailed in Table~\ref{tab:task_taxonomy}.
The \textbf{generation} task encompasses 15 distinct domains:
``E-commerce \& Fintech'', ``Enterprise \& Productivity'', ``Social \& Communication'', ``Data Science \& Analytics'', ``Content Creation \& Multimedia'', ``Entertainment \& Streaming'', ``Game Development \& Gaming'', ``Education \& Learning'', ``Simulation \& Scientific Modeling'', ``Infrastructure \& System Management'', ``DevTools \& Engineering'', ``Logic \& Workflow Visualization'', ``Location Services \& Transit'', ``Information \& Personal Branding'', and ``Lifestyle \& Niche Utilities''.
The \textbf{editing} task consists of sixteen operation types: Data Table, Rich Text Editor, Drag \& Drop Interface, Tree View, Real-time Dashboard, Infinite Scroll, Async Form Validation, File Upload with Progress, Parallax Scrolling, Page Transitions, Particle Effects, Skeleton Loading, Shopping Cart, User Authentication, Multi-step Wizard, and Notification Center.
The \textbf{repair} task addresses eleven types of front-end defects spanning visual, semantic, and interactive dimensions: Occlusion, Crowding, Text Overlap, Alignment, Color \& Contrast, Overflow, Sizing/Proportion, Loss of Interactivity, Semantic Error, Nesting Error, and Missing Attributes.

Our benchmark comprises a total of 1526 tasks, distributed as follows: 123 for Text-Guided Generation, 109 for Vision-Guided Generation, 94 for Video-Guided Generation, 300 for Text-Guided Editing, 300 for Vision-Guided Editing, 300 for Diagnostic Repair, and 300 for Visual-Diagnostic Repair. Each task is annotated with a difficulty level (Easy, Medium, or Hard) based on the complexity of the required functionality, the number of interactive components, and the sophistication of the visual design. A detailed breakdown of per-category counts is provided in Figure~\ref{fig:difficulty_distribution}.

\subsection{Task Type Descriptions}
\label{subsec:task_type_descriptions}

To comprehensively evaluate models across a wide spectrum of real-world web development challenges, WebCompass defines 15 generation application domains, 16 diverse editing task types, and 11 repair defect types. Table~\ref{tab:task_taxonomy} provides an overview, and the following subsections detail each editing and repair task type.

\begin{table*}[t!]
    \centering
    \caption{Detailed taxonomy of Generation, Editing, and Repair tasks in WebCompass. Generation covers 15 application domains; Editing defines 16 modification operations; Repair addresses 11 front-end defect types spanning visual, semantic, and interactive dimensions.}
    \label{tab:task_taxonomy}
    \resizebox{\textwidth}{!}{%
    \renewcommand{\arraystretch}{1.35}
    \begin{tabular}{@{\hspace{6pt}}c l @{\hspace{18pt}} c l @{\hspace{18pt}} c l@{\hspace{6pt}}}
        \toprule
        \multicolumn{2}{c}{\cellcolor{genhead}\textcolor{white}{\textbf{Generation (15 Types)}}}
        & \multicolumn{2}{c}{\cellcolor{edithead}\textcolor{white}{\textbf{Editing (16 Types)}}}
        & \multicolumn{2}{c}{\cellcolor{repairhead}\textcolor{white}{\textbf{Repair (11 Types)}}} \\
        \midrule
        \cellcolor{genbg}\textbf{1} & E-commerce \& Fintech
        & \cellcolor{editbg}\textbf{1} & Data Table
        & \cellcolor{repairbg}\textbf{1} & Occlusion \\
        \rowcolor{rowalt}
        \cellcolor{genbg}\textbf{2} & Enterprise \& Productivity
        & \cellcolor{editbg}\textbf{2} & Rich Text Editor
        & \cellcolor{repairbg}\textbf{2} & Crowding \\
        \cellcolor{genbg}\textbf{3} & Social \& Communication
        & \cellcolor{editbg}\textbf{3} & Drag \& Drop Interface
        & \cellcolor{repairbg}\textbf{3} & Text Overlap \\
        \rowcolor{rowalt}
        \cellcolor{genbg}\textbf{4} & Data Science \& Analytics
        & \cellcolor{editbg}\textbf{4} & Tree View
        & \cellcolor{repairbg}\textbf{4} & Alignment \\
        \cellcolor{genbg}\textbf{5} & Content Creation \& Multimedia
        & \cellcolor{editbg}\textbf{5} & Real-time Dashboard
        & \cellcolor{repairbg}\textbf{5} & Color \& Contrast \\
        \rowcolor{rowalt}
        \cellcolor{genbg}\textbf{6} & Entertainment \& Streaming
        & \cellcolor{editbg}\textbf{6} & Infinite Scroll
        & \cellcolor{repairbg}\textbf{6} & Overflow \\
        \cellcolor{genbg}\textbf{7} & Game Development \& Gaming
        & \cellcolor{editbg}\textbf{7} & Async Form Validation
        & \cellcolor{repairbg}\textbf{7} & Sizing/Proportion \\
        \rowcolor{rowalt}
        \cellcolor{genbg}\textbf{8} & Education \& Learning
        & \cellcolor{editbg}\textbf{8} & File Upload with Progress
        & \cellcolor{repairbg}\textbf{8} & Loss of Interactivity \\
        \cellcolor{genbg}\textbf{9} & Simulation \& Scientific Modeling
        & \cellcolor{editbg}\textbf{9} & Parallax Scrolling
        & \cellcolor{repairbg}\textbf{9} & Semantic Error \\
        \rowcolor{rowalt}
        \cellcolor{genbg}\textbf{10} & Infrastructure \& System Mgmt.
        & \cellcolor{editbg}\textbf{10} & Page Transitions
        & \cellcolor{repairbg}\textbf{10} &  Nesting Error \\
        \cellcolor{genbg}\textbf{11} & DevTools \& Engineering
        & \cellcolor{editbg}\textbf{11} & Particle Effects
        & \cellcolor{repairbg}\textbf{11} & Missing Attributes \\
        \rowcolor{rowalt}
        \cellcolor{genbg}\textbf{12} & Logic \& Workflow Visualization
        & \cellcolor{editbg}\textbf{12} & Skeleton Loading
        & & \\
        \cellcolor{genbg}\textbf{13} & Location Services \& Transit
        & \cellcolor{editbg}\textbf{13} & Shopping Cart
        & & \\
        \rowcolor{rowalt}
        \cellcolor{genbg}\textbf{14} & Information \& Personal Branding
        & \cellcolor{editbg}\textbf{14} & User Authentication
        & & \\
        \cellcolor{genbg}\textbf{15} & Lifestyle \& Niche Utilities
        & \cellcolor{editbg}\textbf{15} & Multi-step Wizard
        & & \\
        \rowcolor{rowalt}
        &
        & \cellcolor{editbg}\textbf{16} & Notification Center
        & & \\
        \bottomrule
    \end{tabular}
    }
\end{table*}

\subsubsection{Editing Task Types}
\label{sec:edit_task_types}

The 16 editing task types span from low-level UI components to full business workflows, ensuring broad coverage of frontend engineering skills. They are organized into four categories:

\paragraph{Complex Components.}
This category includes \textit{Data Table} (sortable, paginated, filterable table with row selection and inline editing), \textit{Rich Text Editor} (WYSIWYG editor with formatting toolbar, link/image insertion, and form-synced output), \textit{Drag \& Drop Interface} (draggable items with drop-zone feedback, cross-container reordering, and state persistence), and \textit{Tree View} (nested expand/collapse tree with cascading selection and search filtering).

\paragraph{Frontend--Backend Integration.}
This category covers \textit{Real-time Dashboard} (live-updating metric cards with animated counters and sparkline charts), \textit{Infinite Scroll} (scroll-triggered lazy loading with skeleton placeholders and end-of-content handling), \textit{Async Form Validation} (debounced server-side validation with inline status indicators and submit gating), and \textit{File Upload with Progress} (drag-and-drop upload with per-file progress bars, queue management, and cancel support).

\paragraph{Advanced Animations.}
This category encompasses \textit{Parallax Scrolling} (multi-layer differential scroll speeds with viewport-triggered fade/scale effects), \textit{Page Transitions} (coordinated enter/exit animations such as fade, slide, and zoom between SPA content views), \textit{Particle Effects} (canvas-based particle system with physics, cursor interaction, and connection lines), and \textit{Skeleton Loading} (shimmer-animated placeholders matching content structure with smooth reveal).

\paragraph{Business Scenarios.}
This category includes \textit{Shopping Cart} (full cart flow with quantity controls, real-time totals, and localStorage persistence), \textit{User Authentication} (login, registration, and password-recovery forms with validation and auth state management), \textit{Multi-step Wizard} (step indicator with per-step validation, cross-step data persistence, and review summary), and \textit{Notification Center} (notification dropdown with unread badge, categorized alerts, and mark-as-read actions).

\subsubsection{Repair Defect Types}
\label{sec:repair_defect_types}

The 11 repair defect types cover visual, semantic, and interactive failure modes commonly encountered in frontend development, organized into three dimensions:

\paragraph{Visual Layout.}
This dimension includes seven defect types: \textit{Occlusion} (one element covers another due to incorrect z-index stacking), \textit{Crowding} (spacing between elements is removed or collapsed, causing visual clutter), \textit{Text Overlap} (text overflows its container and overlaps with adjacent content), \textit{Alignment} (elements are offset from their expected grid or sibling alignment), \textit{Color \& Contrast} (text color is too close to the background, reducing readability), \textit{Overflow} (content exceeds a fixed-size container without proper overflow handling), and \textit{Sizing/Proportion} (elements are given extreme or distorted dimensions).

\paragraph{Semantic Correctness.}
This dimension includes \textit{Semantic Error} (semantic HTML tags replaced with non-semantic equivalents, e.g., \texttt{<h1>} replaced by \texttt{<div>}) and \textit{Nesting Error} (invalid HTML nesting, e.g., \texttt{<a>} inside \texttt{<a>}, or \texttt{<div>} inside \texttt{<p>}).

\paragraph{Interactive Usability.}
This dimension includes \textit{Loss of Interactivity} (interactive elements disabled or blocked via \texttt{pointer-events:~none}) and \textit{Missing Attributes} (accessibility or functional attributes removed, e.g., \texttt{alt}, \texttt{aria-label}).

\section{Evaluation Methodology}
\label{sec:evaluation_methodology}

\begin{figure}[htbp]
    \centering
    \includegraphics[width=1\linewidth]{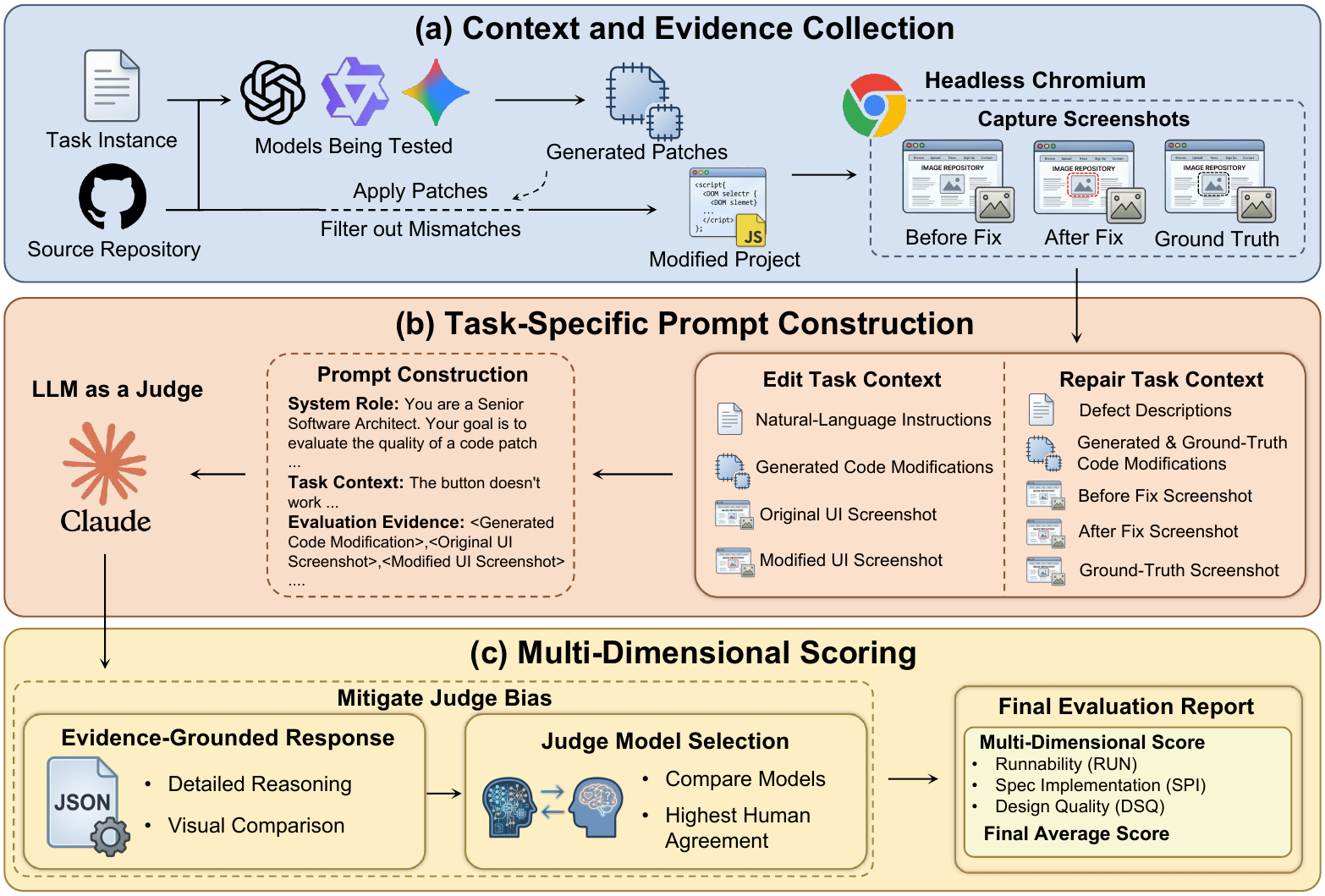}
    \caption{Illustration of the LLM-as-a-Judge evaluation pipeline.}
    \label{fig:llm_as_judge}
\end{figure}

We adopt task-specific evaluation paradigms tailored to the output characteristics of each task family. For \textbf{Editing \& Repair}, where models produce localized code patches, we use \emph{LLM-as-a-Judge} (\S\ref{subsec:llm_judge}). For \textbf{Generation}, where correctness depends on end-to-end runtime behavior, we use \emph{Agent-as-a-Judge} (\S\ref{subsec:agent_judge}). Both paradigms score along three axes---\emph{executability}, \emph{functional}, and \emph{visual}---whose operationalization is task-dependent.
For \textbf{Generation}: \emph{Runnability} (build and launch success), \emph{Spec Implementation} (functional behavior matches the design document), and \emph{Design Quality} (visual polish).
For \textbf{Editing}: \emph{Instruction Targeting} (patch applies and targets the instruction's required locations), \emph{Feature Integrity} (original interactions preserved and new components functional), and \emph{Style Conformance} (visual edit landed and unchanged regions consistent).
For \textbf{Repair}: \emph{Root-Cause Targeting} (patch applies and targets the defect's root cause), \emph{Interaction Integrity} (interactions preserved and interactive-class defects repaired), and \emph{Reference Fidelity} (visual match to the ground-truth fixed screenshot).
We select the judge model based on highest agreement with human annotations (\S\ref{subsec:judge_selection}).

\subsection{LLM-as-a-Judge for Editing \& Repair}
\label{subsec:llm_judge}

For each instance, we apply the predicted patches to the source repository, discard blocks that fail to apply, and launch the modified project in a headless Chromium browser to capture screenshots (Figure~\ref{fig:llm_as_judge}).
The judge receives the original task requirement, the source repository, the model-generated patch, build and runtime logs after patch application, and before/after screenshots captured in headless Chromium. For \textbf{Repair} tasks, it additionally receives the defect description, the ground-truth modifications, and the reference fixed screenshot. It scores checklist items independently along the three task-specific dimensions (0--10 each), produces evidence-grounded structured JSON output, and aggregates the resulting dimension-wise scores into the final task score.

\subsection{Agent-as-a-Judge for Generation}
\label{subsec:agent_judge}

\begin{figure}[htbp]
    \centering
    \includegraphics[width=\linewidth]{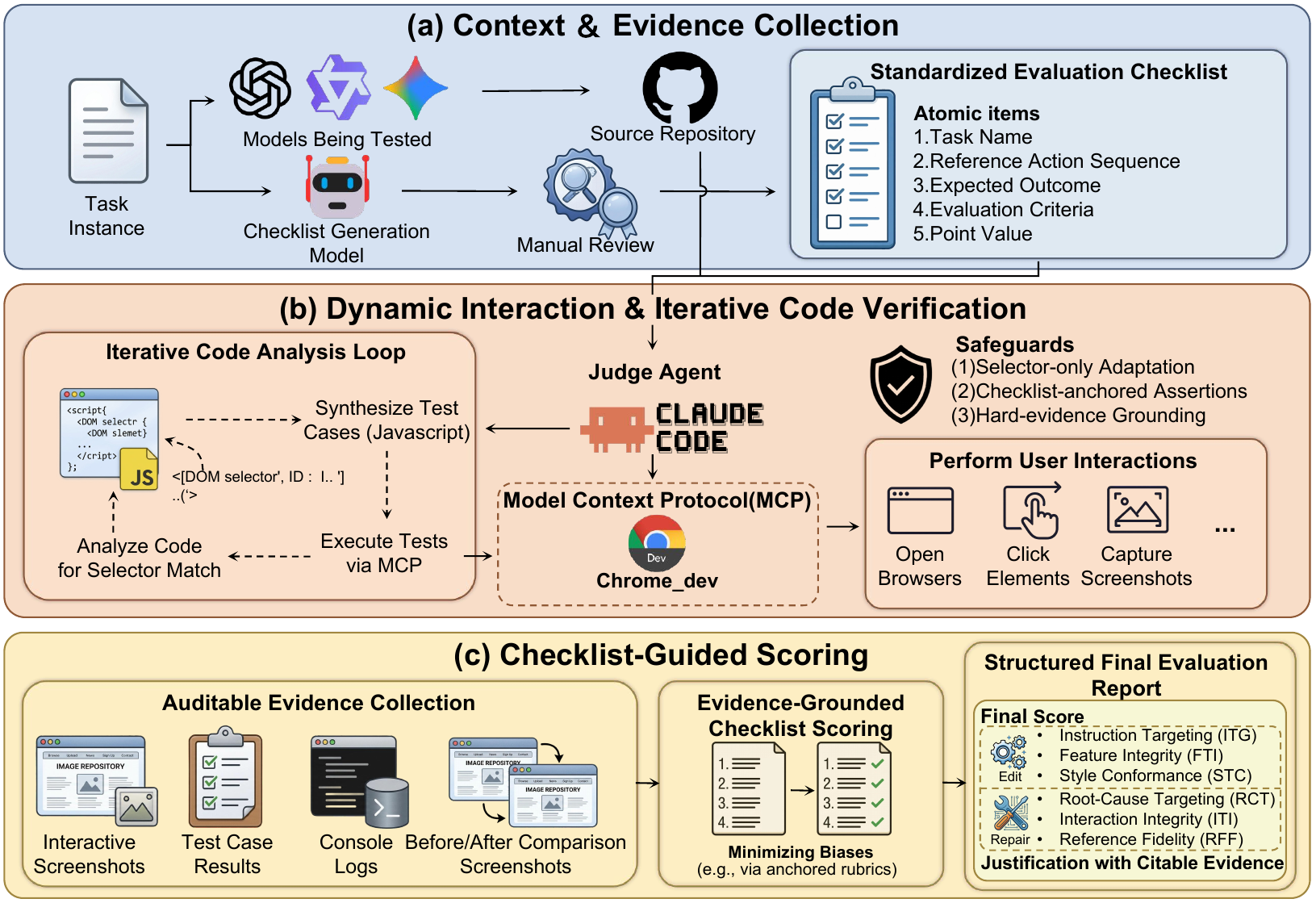}
    \caption{Agent-as-a-Judge evaluation pipeline. The MCP bridge enables bidirectional communication: the agent sends interaction commands to the browser and receives DOM snapshots, console logs, and screenshots as evidence.}
    \label{fig:agent_judge_pipeline}
\end{figure}

Traditional evaluation approaches for web generation fall into two camps, each with a critical blind spot. Pure test-based methods (e.g., unit tests or DOM assertions) can programmatically verify functional correctness—whether a button triggers the right callback or a form validates inputs—but cannot assess visual fidelity, layout harmony, or aesthetic quality. Conversely, screenshot-based comparison methods can capture visual appearance but struggle to verify multi-step interactive behaviors, state transitions, and dynamic content that only manifest through real user interaction. Human acceptance testing naturally combines both capabilities: a QA engineer can inspect the UI visually \emph{and} write ad-hoc test scripts to probe edge cases, switching fluidly between the two modes. To approximate this dual capability in an automated setting, we adopt \textbf{Claude Code} as the evaluation orchestrator paired with the \textbf{Model Context Protocol (MCP)} for browser control. This architecture is deliberately chosen because it endows the judge agent with two complementary verification channels: (1) as a code agent, it can dynamically \emph{synthesize and execute JavaScript test cases} that programmatically inspect DOM states, CSS properties, and functional logic with deterministic precision; and (2) through the MCP bridge to a real browser, it can \emph{simulate authentic user interactions}—clicking, scrolling, typing, navigating—while capturing screenshots and console logs as auditable evidence. Neither channel alone suffices: scripted tests miss visual quality, and browser interaction alone cannot efficiently verify complex state invariants. Their combination enables a unified evaluation loop that closely mirrors how a human tester would accept or reject a web application.

Figure~\ref{fig:agent_judge_pipeline} illustrates our \emph{Agent-as-a-Judge} pipeline. A Code Agent, augmented with the Model Context Protocol (MCP) for real-browser control, evaluates each generated website in four stages:

\textbf{(1)~Checklist generation}: an LLM produces a structured evaluation checklist defining tasks, interaction sequences, expected outcomes, and score values; this checklist remains fixed throughout to prevent circular reasoning.

\textbf{(2)~Browser interaction}: the agent launches the website in headless Chromium, executes checklist interactions (clicking, typing, scrolling, navigation), and records DOM snapshots, console logs, and screenshots as auditable evidence.

\textbf{(3)~Adaptive code verification}: the agent synthesizes executable JavaScript test cases for each checklist item, programmatically verifying DOM states, CSS properties, and functional behaviors. Crucially, when implementation details differ from expectations, the agent adapts only DOM locators (e.g., element selectors and IDs) while keeping all behavioral assertions unchanged---ensuring that evaluation criteria remain anchored to the original specification rather than drifting toward the model's output. Failed tests trigger an iterative debugging loop in which the agent inspects the actual code, diagnoses the mismatch, and re-attempts verification before assigning a score.

\textbf{(4)~Evidence-grounded scoring}: the agent scores each item along Runnability, Spec Implementation, and Design Quality with structured justifications; scores lacking auditable evidence (screenshots, test results, or console logs) are discarded.

Three safeguards prevent evaluation bias: checklist immutability (no new criteria after Stage~1), selector-only adaptation in Stage~3, and mandatory hard-evidence grounding for every score.

All experiments are conducted on a Linux server with per-task execution timeouts to prevent infinite loops or hanging processes. For generation evaluation, we use Claude Code (v2.0.67) as the evaluation orchestrator together with the Chrome DevTools MCP Server (v0.19.0), which provides headless Chromium rendering, DOM inspection, and browser automation capabilities.

\subsection{Scoring and Failure Handling}
\label{subsec:scoring}

\paragraph{Scoring formula.}
For each task instance, let $\{s_1, s_2, \ldots, s_n\}$ denote the individual checklist item scores and $\{m_1, m_2, \ldots, m_n\}$ their corresponding maximum scores. Each item's normalized score is $r_i = s_i / m_i$. To prevent a single zero-scored item from collapsing the entire task score, we apply a smoothing constant $\epsilon = 1$ to any item where $s_i = 0$, replacing its normalized score with $\epsilon / m_i$. The task-level score is then computed as the harmonic mean of the normalized item scores:
\begin{equation}
    s_{\text{task}} = \frac{n}{\sum_{i=1}^{n} \frac{1}{r_i}}
\end{equation}
We choose the harmonic mean over the arithmetic mean because it penalizes imbalanced performance: a web artifact that excels on some criteria while completely failing on others should not receive a high overall score. This score is computed separately for each of the three per-task evaluation dimensions.

\paragraph{Handling cascading failures.}
Web generation tasks frequently produce outputs that fail at different stages of the build--render--interact pipeline, and a na\"{\i}ve application of the scoring formula could yield misleading results. We define explicit fallback strategies for three failure scenarios:
\textbf{1. Complete build failure} (the project does not compile or launch): the functional and visual dimensions are set to $0$; only the executability dimension contributes a meaningful score.
\textbf{2. Partial rendering failure} (the project launches but some pages or components fail to render): the executability dimension is penalized proportionally; the visual dimension is evaluated on the rendered portion (or set to $0$ if nothing is visible); the functional dimension is evaluated only on reachable components.
\textbf{3. Runtime crash} (the project renders initially but crashes during interaction): the executability and visual dimensions are scored on the initial render; the functional dimension is scored only on the testable subset, with untestable items receiving $0$.
These fallback rules ensure that cascading failures degrade scores gracefully rather than producing undefined or inflated results, faithfully reflecting the progressive nature of web application quality.

\section{Experiments}
\label{sec:experiments}

% 本节描述了我们的实验设置和结果概览。我们按任务类型（生成/编辑/修复）和输入模态（文本/图片/视频）组织实验，并报告若干聚焦分析。

This section describes our experimental setup and provides a comprehensive overview of results. We organize experiments by \textbf{task type} (generation / editing / repair) and \textbf{input modality} (text / image / video). Beyond the main results (\S\ref{subsec:main_results}), we report several focused analyses: judge model selection (\S\ref{subsec:judge_selection}), framework-based subset evaluation (\S\ref{subsec:framework_subset}), difficulty-level analysis (\S\ref{subsec:difficulty}), and impact of thinking mode (\S\ref{subsec:think_mode}).

\subsection{Evaluated LLMs and Frameworks}
\label{subsec:models}

We report main benchmark results for ten models from both closed-source and open-source families. All selected models natively support text, image, and video inputs, allowing us to use the same model set across all modalities. Full model details, including auxiliary comparison variants used in later analyses, are provided in Appendix~\ref{model_card}.

We employ \textbf{Claude Code} (v2.0.67) as the evaluation orchestrator and the \textbf{Chrome DevTools MCP Server} (v0.19.0) for browser rendering, DOM inspection, and automated interaction verification in a headless Chromium environment. This setup enables an agent-based evaluation pipeline that programmatically assesses the functional correctness and UI consistency of generated web applications.

\subsection{Main Results}
\label{subsec:main_results}

Table~\ref{tab:main_results} presents the overall and per-task-type scores for all evaluated models.

% ---- 表格 ----
\begin{table*}[htbp]
\centering
\caption{Comparison of models across different task types. \best{Green bold} indicates the best score in each column; \second{blue underline} indicates the second best. Each task has three evaluation dimensions: Generation uses Runnability (RUN), Spec Implementation (SPI), and Design Quality (DSQ); Editing uses Instruction Targeting (ITG), Feature Integrity (FTI), and Style Conformance (STC); Repair uses Root-Cause Targeting (RCT), Interaction Integrity (ITI), and Reference Fidelity (RFF). Overall is the arithmetic mean of all nine dimension scores.}
\label{tab:main_results}
\resizebox{\textwidth}{!}{%
\renewcommand{\arraystretch}{1.25}
\begin{tabular}{@{}l|ccc|ccc|ccc|c@{}}
\toprule
\multirow{2}{*}{\textbf{Models}}
 & \multicolumn{3}{c|}{\textbf{Generation}}
 & \multicolumn{3}{c|}{\textbf{Editing}}
 & \multicolumn{3}{c|}{\textbf{Repair}}
 & \multirow{2}{*}{\textbf{Overall}} \\
\cmidrule(l){2-4} \cmidrule(l){5-7} \cmidrule(l){8-10}
 & \textbf{RUN.} & \textbf{SPI.} & \textbf{DSQ.}
 & \textbf{ITG.} & \textbf{FTI.} & \textbf{STC.}
 & \textbf{RCT.} & \textbf{ITI.} & \textbf{RFF.}
 & \\
\midrule

% ---- Closed-Source ----
\rowcolor{closedtag}
\multicolumn{11}{c}{\textit{\textcolor{headerblue}{\textbf{Closed-Source Large Language Models}}}} \\
\midrule
\rowcolor{rowgray}  Claude-Opus-4.5
    & \best{77.18} & \best{68.95} & 62.26
    & \best{71.86} & \best{65.82} & \best{60.83}
    & 48.45 & 85.54 & 65.71
    & \best{67.40} \\
    Gemini-3-Pro-Preview
    & 74.05 & 55.76 & \best{64.07}
    & \second{69.52} & \second{65.14} & \second{58.16}
    & \best{54.16} & \best{87.30} & \best{72.00}
    & \second{66.68} \\
\rowcolor{rowgray}    Gemini-3-Flash-Preview
    & \second{74.87} & 54.32 & \second{62.42}
    & 65.95 & 62.35 & 57.21
    & \second{53.18} & \second{86.84} & \second{71.65}
    & 65.42 \\
    GPT-5.2
    & 75.38 & \second{60.22} & 55.92
    & 66.97 & 62.70 & 56.63
    & 41.24 & 79.33 & 58.70
    & 61.90 \\
\rowcolor{rowgray}     Claude-Sonnet-4.5
    & 65.30 & 50.37 & 56.78
    & 60.06 & 53.71 & 45.51
    & 40.44 & 80.63 & 61.31
    & 57.12 \\
\midrule

% ---- Open-Source ----
\rowcolor{opentag}
\multicolumn{11}{c}{\textit{\textcolor{headerblue}{\textbf{Qwen3-VL Series Open-Source Large Language Models}}}} \\
\midrule
\rowcolor{rowgray} 235B-A22B-Instruct
    & \second{61.26} & \best{42.14} & \best{47.06}
    & \best{27.74} & \second{25.48} & \best{23.53}
    & \best{27.30} & \best{68.87} & \best{46.88}
    & \best{41.14} \\
                  235B-A22B-Thinking
    & \best{63.86} & \second{35.02} & \second{45.21}
    & 22.15 & 21.67 & 19.06
    & \second{27.02} & \second{68.74} & \second{46.28}
    & \second{38.78} \\
\rowcolor{rowgray} 32B-Instruct
    & 50.39 & 25.62 & 34.56
    & \second{26.96} & \best{26.62} & \second{22.78}
    & 24.67 & 61.93 & 43.27
    & 35.20 \\
30B-A3B-Thinking
    & 47.37 & 20.87 & 37.47
    & 19.82 & 21.21 & 18.20
    & 18.08 & 51.85 & 31.31
    & 29.58 \\
\rowcolor{rowgray} 30B-A3B-Instruct
    & 41.79 & 20.80 & 29.28
    & 20.57 & 20.97 & 17.93
    & 19.32 & 50.71 & 31.35
    & 28.08 \\

\bottomrule
\end{tabular}%
}

\end{table*}

Several key patterns emerge.

\textbf{Model ranking and the closed--open gap.}
Claude-Opus-4.5 and Gemini-3-Pro-Preview achieve the highest Overall scores (67.40 and 66.68, respectively) with complementary strengths: Claude leads Generation RUN (77.18) and Editing ITG (71.86), while Gemini leads Repair RCT (54.16) and Repair RFF (72.00). The closed--open gap is substantial: the best open-source model (Qwen3-VL-235B-A22B-Instruct) reaches an Overall of 41.14, trailing the top closed-source model by over 26 points. Smaller open models (30B variants) fall further, reaching less than half the top closed-source scores.

\textbf{Task-type patterns.}
For closed-source models, Generation and Editing consistently follow the ordering executability $>$ functional $>$ visual (e.g., Claude-Opus-4.5: RUN 77.18 $>$ SPI 68.95 $>$ DSQ 62.26 on Generation). Repair shows a different pattern: ITI $\gg$ RFF $>$ RCT (e.g., Gemini-3-Pro-Preview: 87.30 $>$ 72.00 $>$ 54.16). This ordering is explained by the task structure: Interaction Integrity trends high because 9 of 11 defect types are visual or semantic---the interactive layer is rarely affected, so preservation is nearly automatic; the 2 interactive-class defects (Loss of Interactivity, Missing Attributes) are localized enough that models can usually repair them. Reference Fidelity is mid-range because matching the gold fixed screenshot is nontrivial. Root-Cause Targeting is lowest because correctly locating the defect's root cause without introducing new errors remains the hardest part of repair. Note that the functional and visual axes measure different capabilities across tasks: Editing's Feature Integrity tests both preservation and new-component functionality, whereas Repair's Interaction Integrity is primarily regression safety; similarly, Editing's Style Conformance evaluates edit outcome fidelity, while Repair's Reference Fidelity measures closeness to a gold reference. Editing is especially challenging for open-source models, where scores fall to 18--28 across dimensions, revealing a major gap in context-aware code modification.

\textbf{Visual quality as the persistent bottleneck.}
Across all ten models, the visual dimension is the lowest-scoring axis in Generation and Editing (Design Quality and Style Conformance, respectively). Even Gemini-3-Pro-Preview, the strongest model on this axis, reaches only 64.07 on Generation DSQ. The gap is wider for weaker models and consistent across task types. Notably, Gemini-3-Pro-Preview and Gemini-3-Flash-Preview outperform GPT-5.2 on the visual axis despite comparable executability scores, indicating that visual fidelity and functional correctness do not scale in lockstep.

\subsection{Further Analysis}
\label{subsec:further_analysis}

\subsubsection{Judge Model Selection}
\label{subsec:judge_selection}
\begin{figure*}[t]
  \centering
  \includegraphics[width=\textwidth]{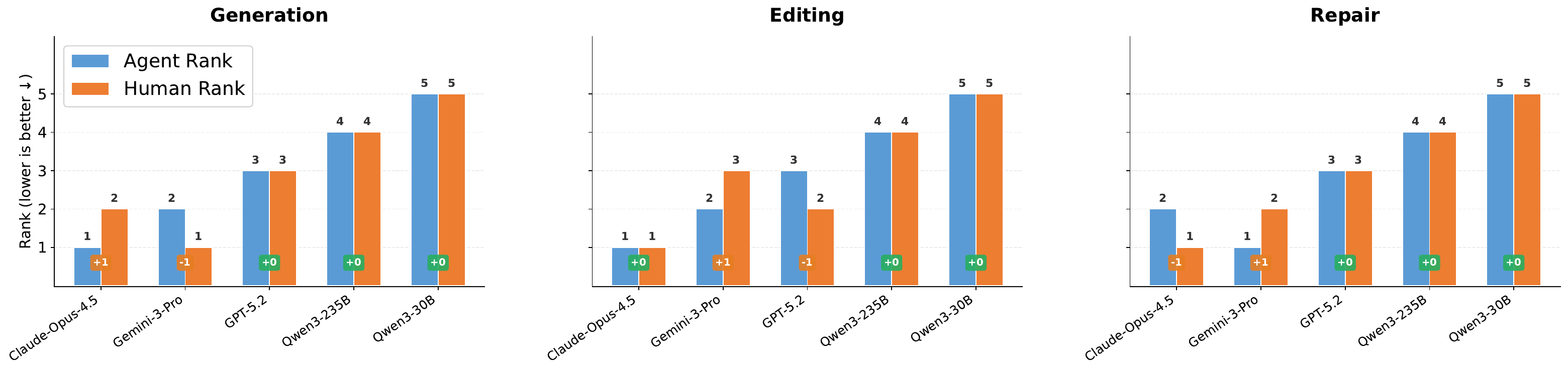}
  \caption{Comparison of model rankings between agent-based automatic evaluation and human evaluation across three tasks. In most cases, the rank difference is zero or at most one, indicating strong agreement between the automatic evaluator and human annotators.}
  \label{fig:rank_comparison}
\end{figure*}

To validate automated evaluation reliability, we compare three Claude-family judge models (Opus-4.5, Sonnet-4.5, Haiku-4.5) against human judgments on a 200-sample subset. As shown in Table~\ref{tab:judge_selection}, Claude-Opus-4.5 achieves the highest human agreement (Pearson $r$ of 0.93--0.96 across tasks), and is adopted as the default judge. Notably, all judges show higher agreement on edit/repair tasks than on generation, consistent with the more constrained solution space of patch-based tasks. As shown in Figure~\ref{fig:rank_comparison}, a comparison of full model rankings between the agent-based evaluator and human annotators further confirms strong alignment, with most rank differences being zero or at most one, validating the automatic evaluation protocol as a reliable proxy for human judgment.

\begin{table}[ht]
\centering
\caption{Judge model comparison. We report human agreement (Pearson $r$) and estimated cost per sample. \best{Green bold}: best; \second{blue underline}: second best.}
\label{tab:judge_selection}
% \resizebox{\columnwidth}{!}{
\renewcommand{\arraystretch}{1.0}
\begin{tabular}{@{}l|c|c|c|c@{}}
\toprule
\multirow{2}{*}{\textbf{\textcolor{headerblue}{Judge Model}}}
 & \textbf{\textcolor{headerblue}{Generation}}
 & \textbf{\textcolor{headerblue}{Editing}}
 & \textbf{\textcolor{headerblue}{Repair}}
 & \textbf{\textcolor{headerblue}{Cost Analysis}} \\
\cmidrule(l){2-2} \cmidrule(l){3-3} \cmidrule(l){4-4} \cmidrule(l){5-5}
 & \textbf{\textcolor{headerblue}{$r$}}
 & \textbf{\textcolor{headerblue}{$r$}}
 & \textbf{\textcolor{headerblue}{$r$}}
 & \textbf{\textcolor{headerblue}{Cost}} \\
\midrule
Claude-Opus-4.5 & \best{0.93} & \best{0.94} & \best{0.96} & \$4.66 \\
\rowcolor{rowgray}
Claude-Sonnet-4.5 & \second{0.88} & \second{0.90} & \second{0.89} & \$2.34 \\
Claude-Haiku-4.5 & 0.76 & 0.79 & 0.81 & \$1.02 \\
\bottomrule
\end{tabular}%
% }
\end{table}

This comparison also reveals a clear cost--quality trade-off. The cost column in Table~\ref{tab:judge_selection} reports the average API token expenditure (in USD) for evaluating a single task instance. Claude-Sonnet-4.5 is cheaper but consistently trails Opus in agreement, while Haiku shows a substantial drop in alignment despite the lowest cost. We therefore choose Opus-4.5 as the default judge because judge reliability is foundational to benchmark validity, and the additional evaluation cost is justified by the stronger agreement with human assessment.

\subsubsection{Subset Evaluation on Different Front-End Frameworks}
\label{subsec:framework_subset}

To assess how framework choice affects model performance, we evaluate four representative models (GPT-5.2, Gemini-3-Pro-Preview, Claude-Opus-4.5, Qwen3-VL-235B-A22B-Instruct) on a 180-task subset (60 per task category), each completed in \textbf{React}, \textbf{Vue}, and \textbf{Vanilla} (plain HTML/CSS/JS). Figure~\ref{fig:framework_bar} presents the overall scores; per-dimension breakdowns are in Appendix~\ref{app:framework_detail}.

\begin{figure*}[t]
    \centering
    \includegraphics[width=\linewidth]{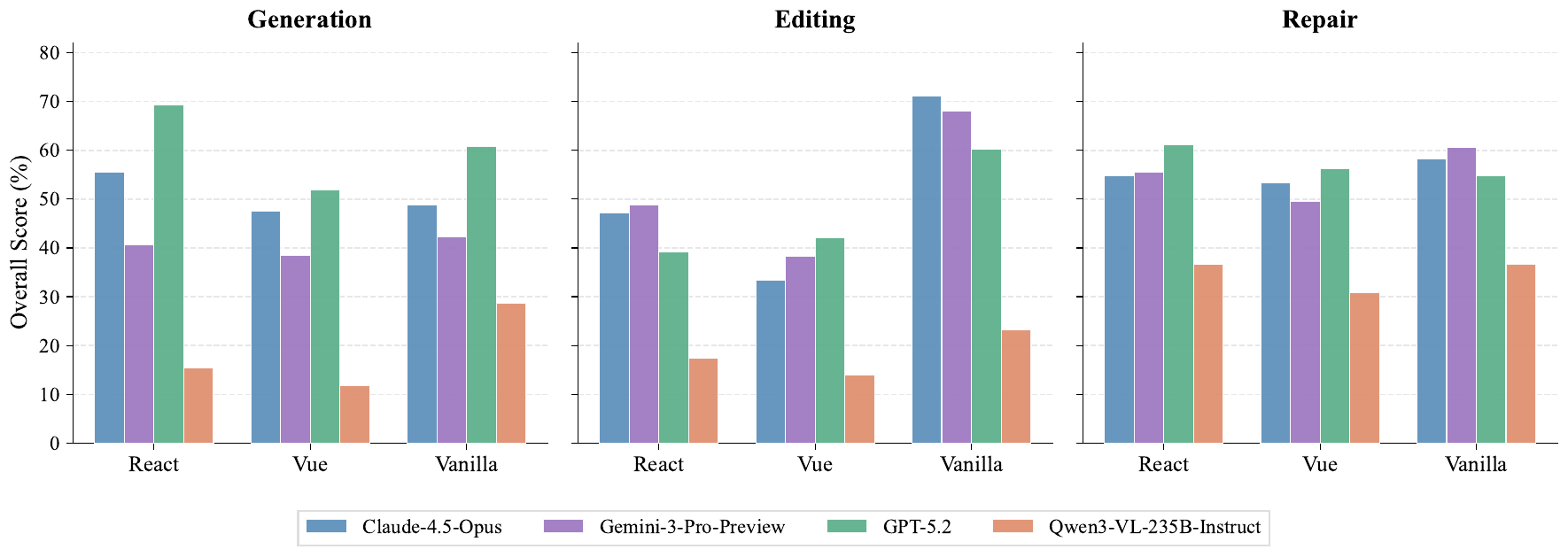}
    \caption{Overall scores across front-end frameworks for four representative models on Generation, Editing, and Repair tasks. Scores are computed as the harmonic mean of the three per-task evaluation dimensions.}
    \label{fig:framework_bar}
\end{figure*}

Three key findings emerge.
\textbf{(1) Vanilla dominates Generation and Editing, but not Repair.} Across all four models, framework-free code consistently yields the highest scores in Generation and Editing. In Repair, however, the Vanilla advantage diminishes: for instance, GPT-5.2 achieves its best Repair score with React. We attribute this to a structural difference between task types: Generation and Editing require producing substantial new code, where Vanilla's absence of build toolchains, framework-specific syntax (e.g., JSX, template directives), and component lifecycle conventions reduces the surface area for errors. Repair, by contrast, demands precise localization and modification of existing defects, where React's explicit component boundaries and unidirectional data flow may help models isolate faulty code regions more effectively than unstructured Vanilla codebases.
\textbf{(2) Vue consistently underperforms.} Vue yields the lowest scores in the majority of model--task combinations. A plausible contributing factor lies in Vue's single-file component (SFC) format, which interleaves three heterogeneous syntax modes---HTML-like templates with custom directives (\texttt{v-if}, \texttt{v-for}, \texttt{@click}), JavaScript/TypeScript logic, and scoped CSS---within a single file. This demands simultaneous coordination across markup, logic, and styling, increasing the likelihood of cross-block inconsistencies. By comparison, React's JSX keeps rendering logic within standard JavaScript, and Vanilla avoids framework abstractions entirely.
\textbf{(3) Open-source models share the same framework sensitivity pattern} (peaking on Vanilla for Generation/Editing) but at a uniformly lower performance ceiling, suggesting that the observed framework preferences are primarily driven by inherent task--framework interactions rather than model-specific factors.

\subsubsection{Task-Type Breakdown}
\label{subsec:task_breakdown}

\begin{figure*}[t]
    \centering
    \includegraphics[width=\linewidth]{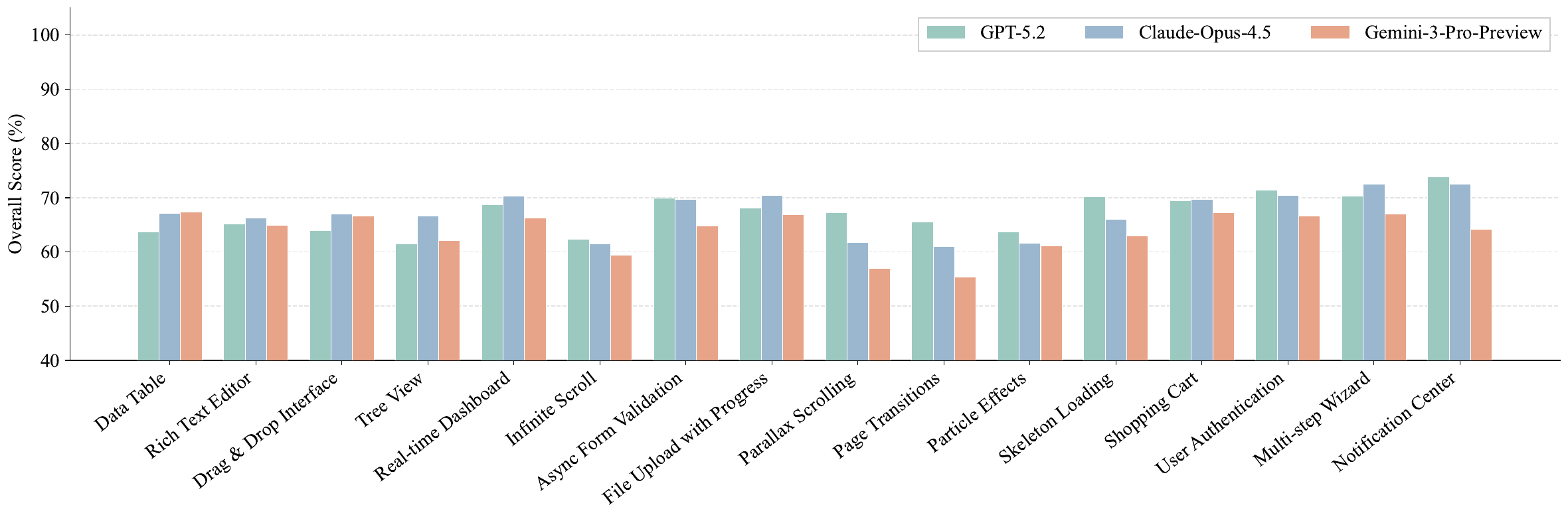}
    \caption{Overall score breakdown for editing tasks across 16 operation types. Scores are computed as the harmonic mean of Instruction Targeting, Feature Integrity, and Style Conformance per subtask, averaged over all instances containing that operation type.}
    \label{fig:edit_subtask_bars}
\end{figure*}

\begin{figure*}[t]
    \centering
    \includegraphics[width=\linewidth]{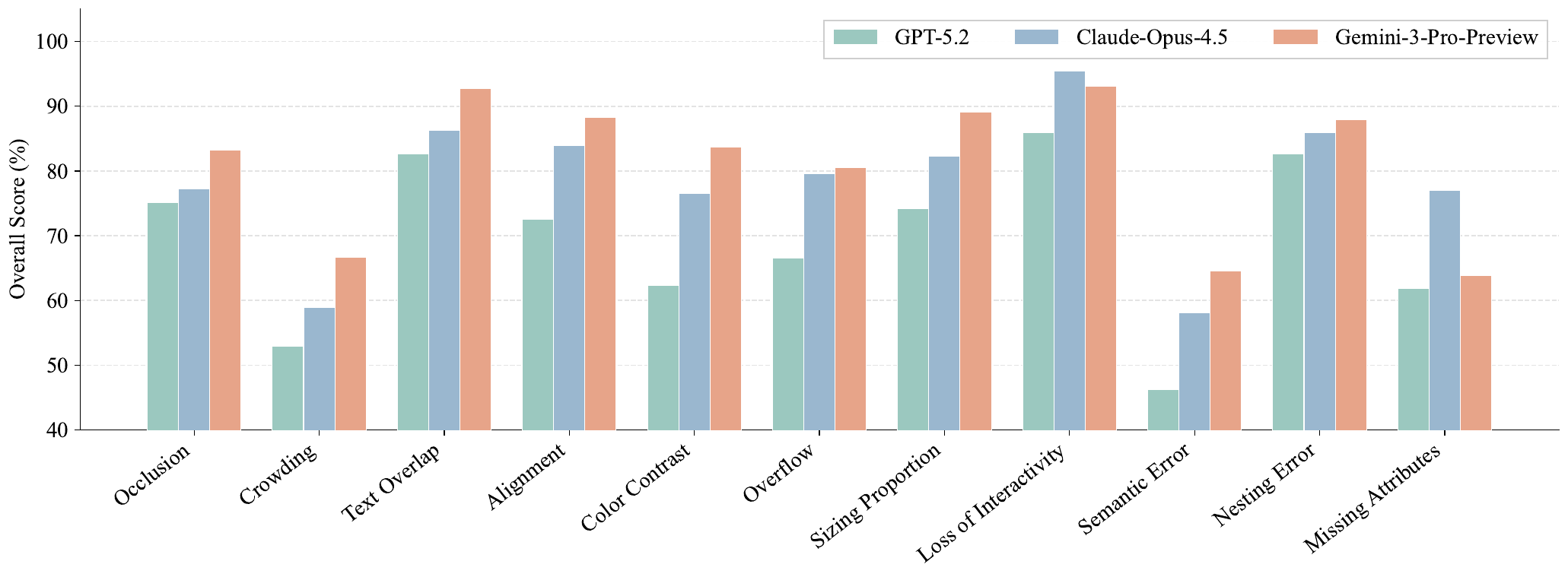}
    \caption{Overall score breakdown for repair tasks across 11 defect categories. Scores are computed identically to Figure~\ref{fig:edit_subtask_bars}.}
    \label{fig:repair_subtask_bars}
\end{figure*}

To reveal where strong models succeed and fail, we further decompose Edit and Repair into fine-grained subtask categories for the three best closed-source models. Figures~\ref{fig:edit_subtask_bars} and~\ref{fig:repair_subtask_bars} report the harmonic-mean score for each subtask type.

\paragraph{Editing: animation-heavy operations are the hardest.}
A clear difficulty hierarchy emerges across editing operation types (Figure~\ref{fig:edit_subtask_bars}): Business Scenario tasks such as Shopping Cart and Multi-step Wizard are consistently the easiest, followed by Real-time \& Async tasks, then Interactive Components, with Advanced Animation tasks such as Parallax Scrolling, Page Transitions, and Particle Effects forming the hardest category. This ordering is stable across all three models, suggesting that editing difficulty scales with the degree of visual dynamism and cross-component coordination required.

\paragraph{Repair: semantic defects remain the main bottleneck.}
A similar difficulty gradient appears in repair (Figure~\ref{fig:repair_subtask_bars}). Structural and interactive defects such as Loss of Interactivity, Nesting Error, and Text Overlap are reliably fixed, as they often manifest in localized DOM structures or event handlers. Semantic-level defects, however, prove substantially harder: Semantic Error elicits the lowest scores across all models, followed by Crowding and Missing Attributes. These categories require reasoning about design intent and implicit visual constraints that go beyond pattern-matching on code structure.

\paragraph{Consistency matters more than isolated wins.}
An instructive discrepancy emerges for editing: GPT-5.2 outperforms Gemini-3-Pro-Preview on 13 of 16 subtask types when averaged per category, yet trails on instance-level scores in Table~\ref{tab:main_results}. This reversal stems from our harmonic-mean aggregation---GPT-5.2 exhibits higher cross-subtask variance, and the harmonic mean penalizes low outliers steeply. This highlights that multi-requirement evaluation rewards consistency, not just peak subtask performance. No such reversal occurs for repair, where Gemini-3-Pro-Preview leads on all 11 defect categories, reflecting genuinely superior repair capability.

\subsubsection{Difficulty-Level Analysis}
\label{subsec:difficulty}

Each WebCompass instance is annotated as \textit{Easy}, \textit{Medium},
or \textit{Hard} according to required functionality, number of interactive
components, and visual sophistication. This stratification lets us examine
how model quality degrades with task complexity.

\begin{figure*}[htbp]
    \centering
    \includegraphics[width=0.95\textwidth]{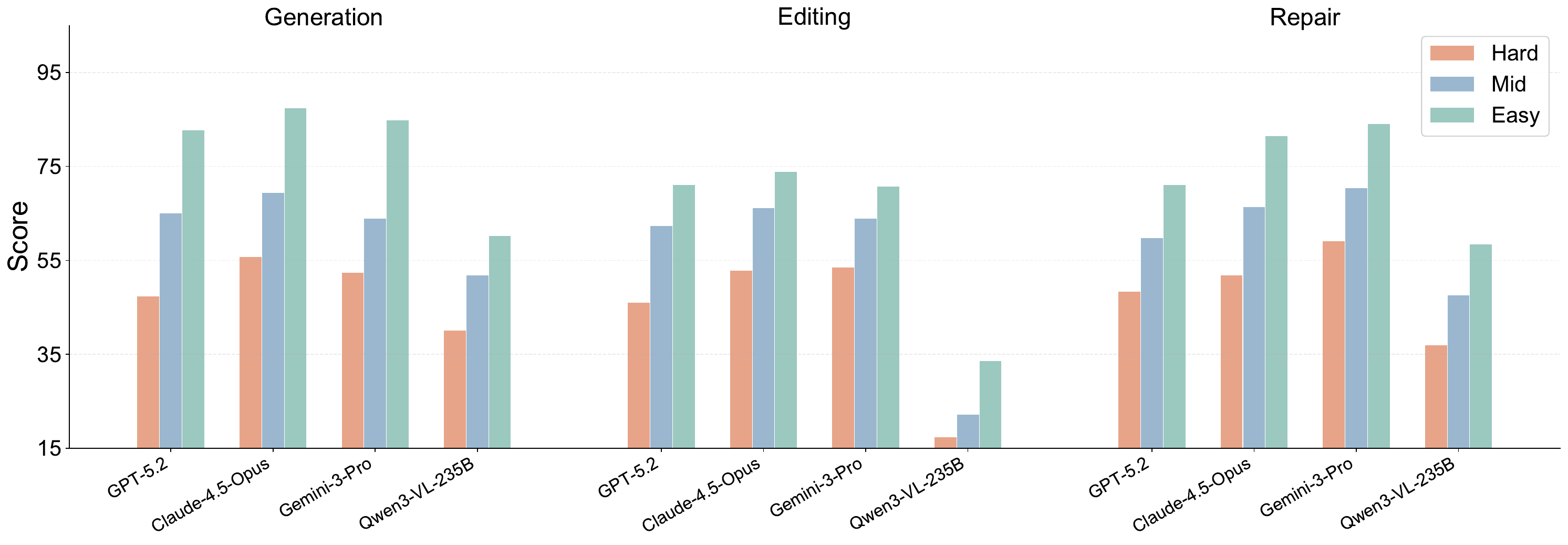}
    \caption{Performance comparison across Generation, Editing, and Repair tasks by difficulty level.}
    \label{fig:overview-difficulty}
\end{figure*}

\begin{figure*}[htbp]
    \centering
    \includegraphics[width=0.95\textwidth]{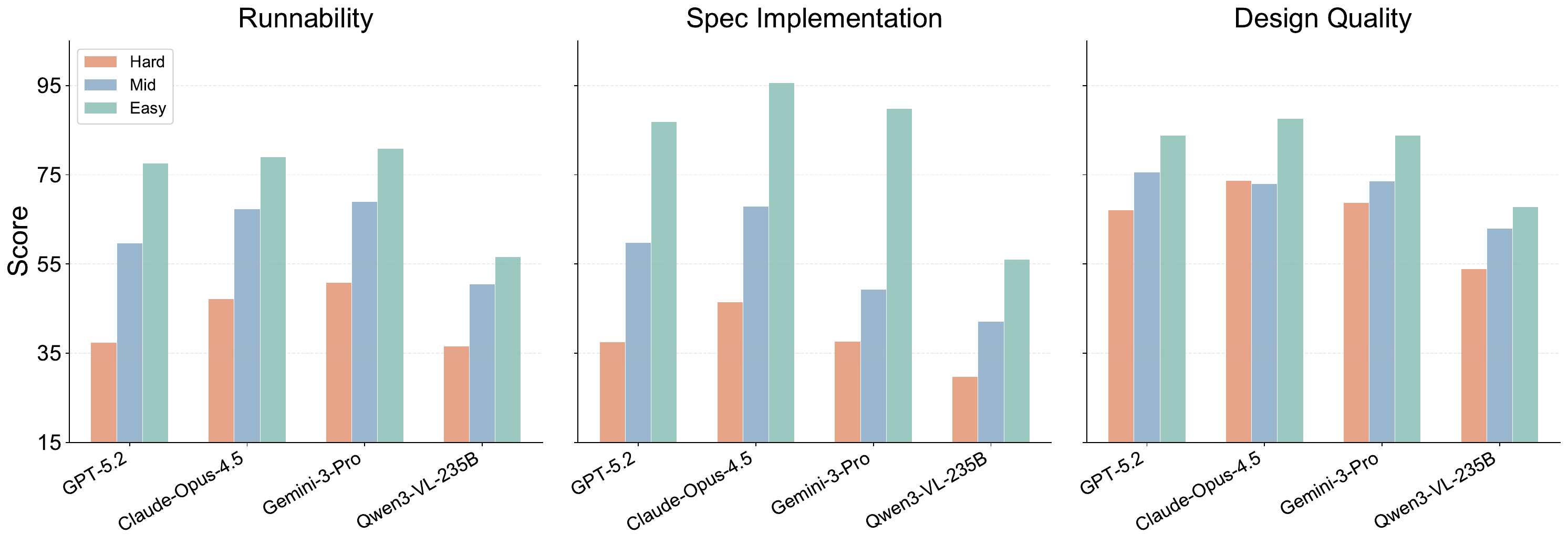}
    \caption{Generation task: per-dimension scores (Runnability, Spec Implementation, and Design Quality) across three difficulty levels. Each model has three bars representing Hard (red), Medium (blue), and Easy (green).}
    \label{fig:gen-difficulty}
\end{figure*}

\begin{figure*}[t]
    \centering
    \includegraphics[width=0.95\textwidth]{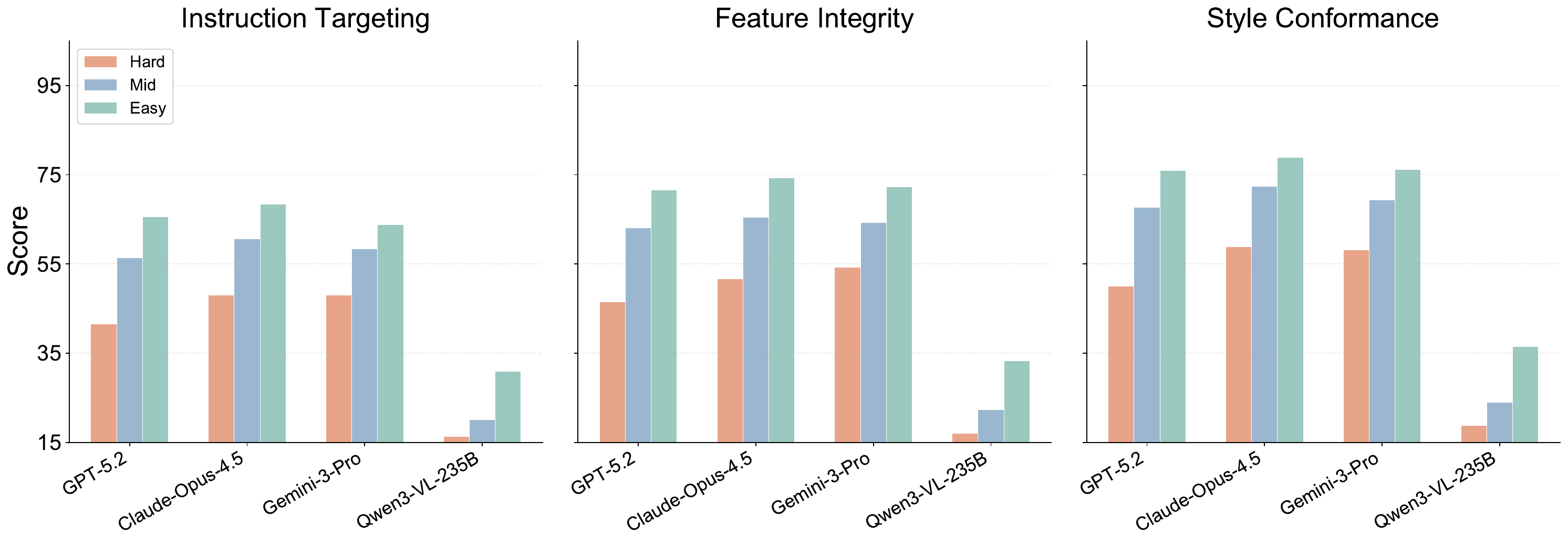}
    \caption{Edit task: per-dimension scores (Instruction Targeting, Feature Integrity, and Style Conformance) across three difficulty levels.}
    \label{fig:edit-difficulty}
\end{figure*}

\begin{figure*}[t]
    \centering
    \includegraphics[width=0.95\textwidth]{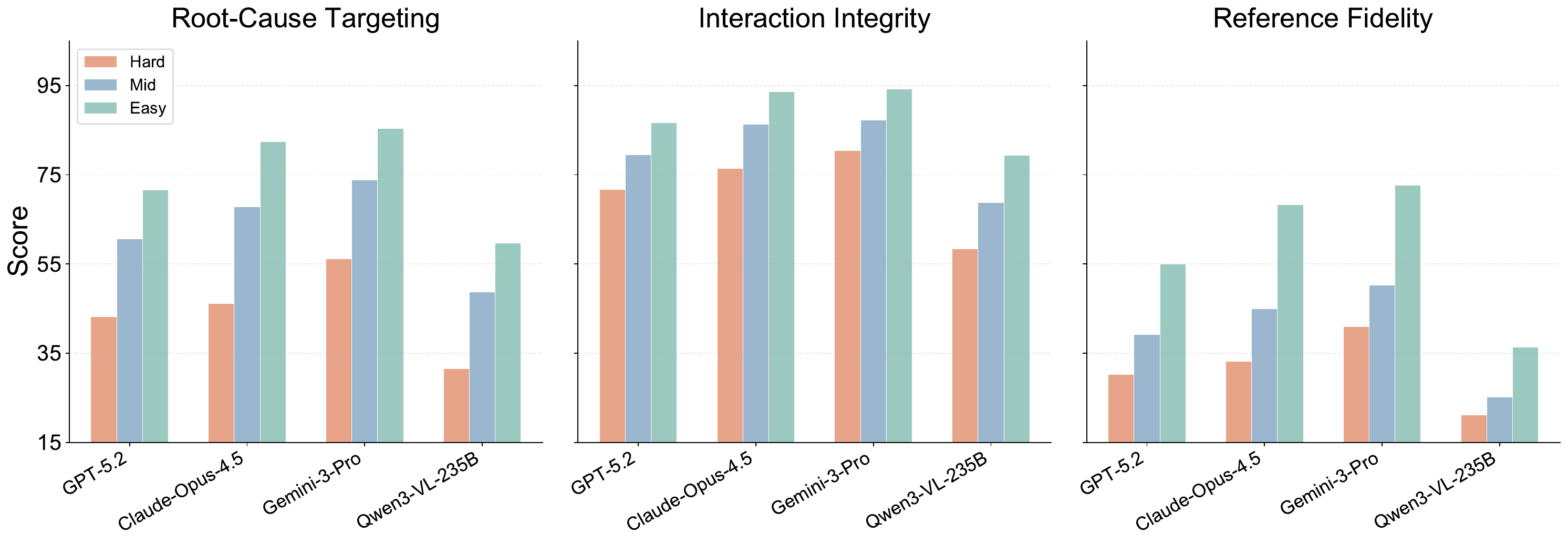}
    \caption{Repair task: per-dimension scores (Root-Cause Targeting, Interaction Integrity, and Reference Fidelity) across three difficulty levels.}
    \label{fig:repair-difficulty}
\end{figure*}

To further investigate how model capabilities scale with task complexity,
we break down the evaluation results across three difficulty levels
(\textit{Easy}, \textit{Medium}, \textit{Hard}) for each task category.
Figure~\ref{fig:overview-difficulty} presents an overview across all
three task families. Across all task families and evaluation dimensions,
model scores consistently decrease as difficulty increases
(Figures~\ref{fig:gen-difficulty},~\ref{fig:edit-difficulty},
and~\ref{fig:repair-difficulty}). This degradation is particularly
striking in generation on the \textit{Spec Implementation} dimension
(Figure~\ref{fig:gen-difficulty}), where Hard tasks require implementing
more complex user flows, multi-step state transitions, and richer dynamic
behavior. For example, Gemini-3-Pro-Preview drops from 89.83 on Easy
generation tasks to 37.64 on Hard ones, suggesting that faithfully
implementing the full functional spec becomes disproportionately
challenging as task complexity grows.

\paragraph{Cross-task observations.}
As shown in Figure~\ref{fig:overview-difficulty}, Qwen3-VL-235B-A22B-Instruct
consistently ranks last across all task--difficulty settings, with its weakest
performance appearing most clearly on Editing. In contrast, GPT-5.2,
Claude-4.5 Opus, and Gemini-2.5 Pro remain substantially stronger across the
board, although their relative advantages vary by task. Overall, performance
drops as difficulty increases for all models, while the relative ranking is
largely preserved, suggesting that harder front-end tasks degrade performance
broadly rather than uniformly widening the gap between models.

\subsubsection{Patch Complexity Analysis}
\label{subsec:patch_complexity}

Beyond output quality, we analyze the structural complexity of model-generated patches using two complementary metrics: \textbf{Changed Lines} (added plus deleted lines) and \textbf{Patch Count} (number of contiguous diff hunks). For repair, we additionally compare against the human-authored ground-truth patches.

\begin{figure*}[t]
    \centering
    \includegraphics[width=0.8\linewidth]{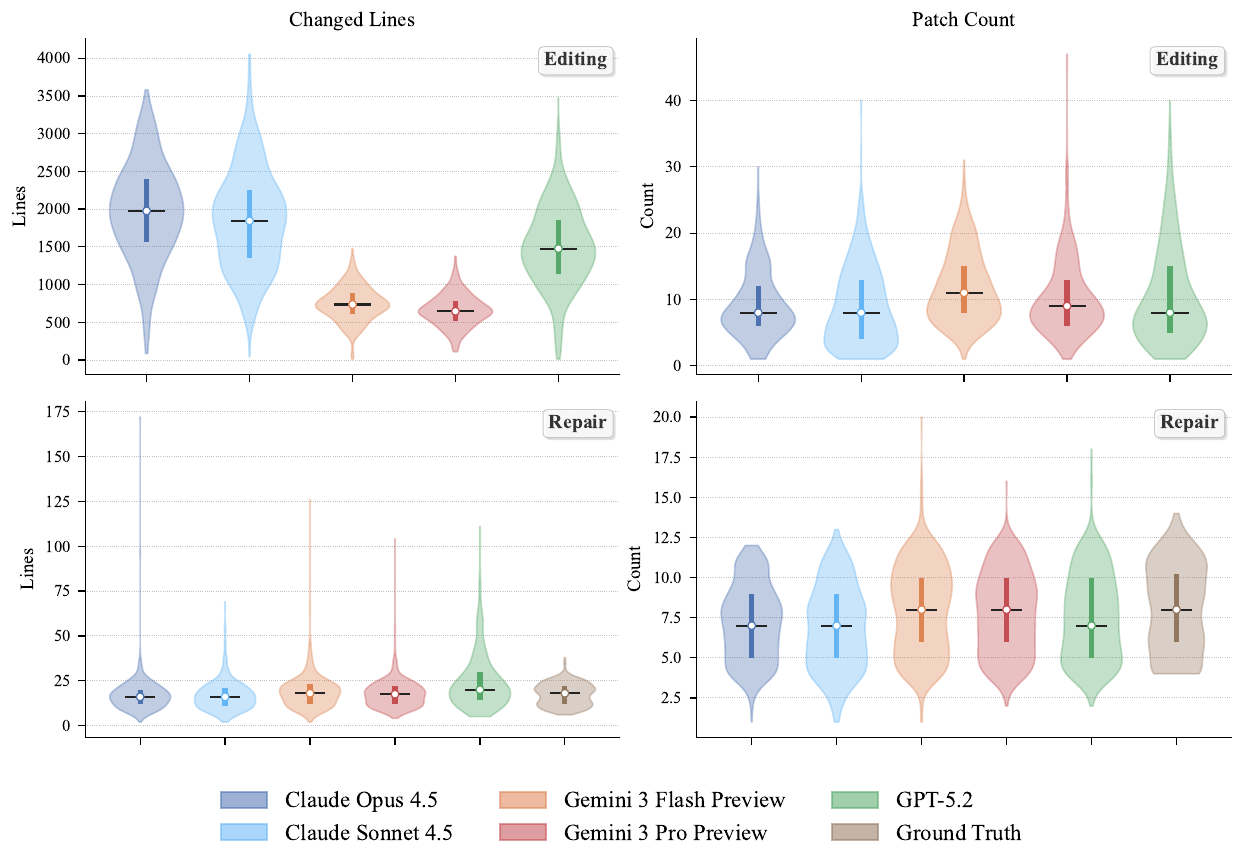}
    \caption{Distribution of patch complexity across models. Top row: Edit tasks; bottom row: Repair tasks (with Ground Truth baseline). Each violin shows the full distribution; the thick bar marks the interquartile range (Q1--Q3) and the white dot marks the median.}
    \label{fig:patch_violin}
\end{figure*}

Figure~\ref{fig:patch_violin} reveals two patterns. First, editing patches are far larger than repair patches, matching the task structure: editing often introduces new components or rewrites existing interaction flows, whereas repair usually targets localized defects. Edit tasks yield median patch sizes of 646--1{,}976 changed lines across models, while repair patches are much smaller, with medians of 16--19 lines. Second, stronger models do not simply generate larger patches. Claude-Opus produces the largest edit patches, roughly three times larger than Gemini models, despite only modest quality differences, while repair patches stay close to the human-authored baseline but exhibit heavier right tails, indicating occasional over-editing. Together these results suggest that successful web coding depends less on patch size itself than on targeting the right code regions with coherent, well-localized updates.

\subsubsection{Stability Analysis: Worst-of-N Evaluation}

\begin{figure}
    \centering
    \includegraphics[width=0.6\linewidth]{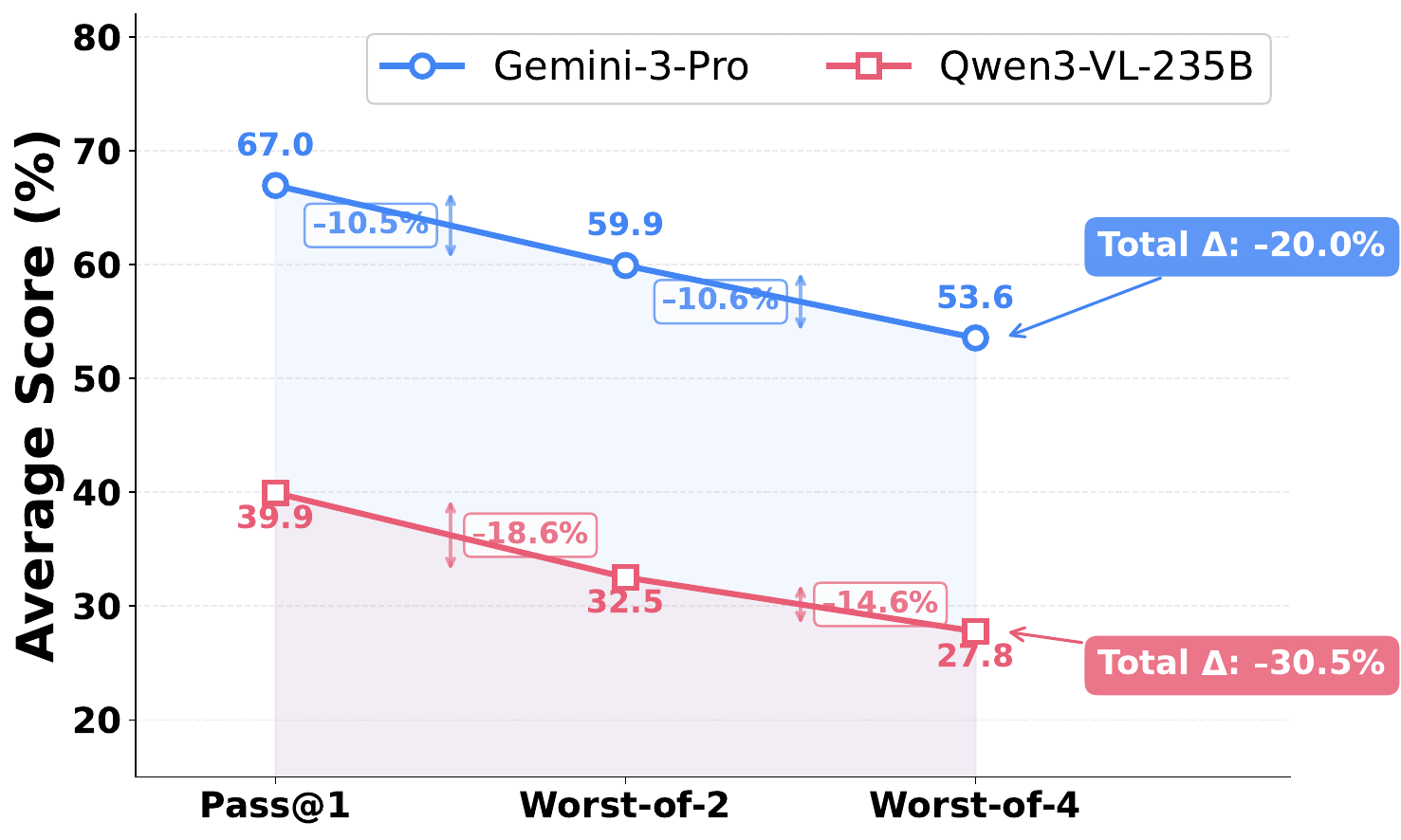}
    \caption{Consistency \& Stability: Score Degradation under Worst-of-N}
    \label{fig:worst_of_n}
\end{figure}

Pass@1 reflects average-case capability but may mask output inconsistency. We adopt the Worst-of-$n$ (W@$n$) protocol, sampling $n=4$ independent generations per task and reporting the minimum score to capture the realistic lower bound of model performance.

As shown in Figure~\ref{fig:worst_of_n}, both models degrade monotonically from Pass@1 to W@4, but at different rates. \textbf{Gemini-3-Pro-Preview} retains $\sim$80\% of its Pass@1 performance at W@4 (66.96$\rightarrow$53.56), with all dimensions remaining above 49\%. \textbf{Qwen3-VL-235B-A22B-Instruct} degrades more sharply, retaining only $\sim$69.5\% (39.95$\rightarrow$27.78), with W@4 scores in the Edit category falling below 16\% on Instruction Targeting—indicating near-complete failure in worst-case scenarios.

These results reveal that Gemini's advantage extends beyond higher average scores to \textit{greater output consistency}. Since users typically rely on a single generation rather than selecting from multiple samples, output stability remains a critical open challenge for frontier models in front-end code generation. 
% Detailed per-task and per-dimension breakdowns are provided in Appendix~\ref{app:worst_of_n}.

\subsubsection{Text-Only vs. Vision-Language Models}
\label{subsec:text_vs_vlm}

To investigate whether visual grounding helps or hurts front-end code generation when the task itself is text-based, we compare \textsc{Qwen3-32B} and \textsc{Qwen3-VL-32B-Instruct} on three representative text-only task types: Text-Guided Generation, Text-Guided Editing, and Diagnostic Repair.

\begin{table*}[t]
\centering
\caption{Comparison of \textsc{Qwen3-32B} and \textsc{Qwen3-VL-32B-Instruct} on the text-only subset across three web-development task types. \best{Green bold} indicates the best score in each column. Dimension abbreviations follow Table~\ref{tab:main_results}; Overall is the arithmetic mean of all nine dimension scores.}
\label{tab:qwen3_web_comparison}
\resizebox{\textwidth}{!}{%
\renewcommand{\arraystretch}{1.25}
\begin{tabular}{@{}l|ccc|ccc|ccc|c@{}}
\toprule
\multirow{2}{*}{\textbf{Models}}
 & \multicolumn{3}{c|}{\textbf{Generation}}
 & \multicolumn{3}{c|}{\textbf{Edit}}
 & \multicolumn{3}{c|}{\textbf{Repair}}
 & \multirow{2}{*}{\textbf{Overall}} \\
\cmidrule(l){2-4} \cmidrule(l){5-7} \cmidrule(l){8-10}
 & \textbf{RUN.} & \textbf{SPI.} & \textbf{DSQ.}
 & \textbf{ITG.} & \textbf{FTI.} & \textbf{STC.}
 & \textbf{RCT.} & \textbf{ITI.} & \textbf{RFF.}
 & \\
\midrule
Qwen3-32B
    & \best{56.28} & 6.50 & 49.10
    & 21.52 & 21.17 & 17.92
    & 21.42 & 56.73 & 33.72
    & 31.60 \\
\midrule
Qwen3-VL-32B-Instruct
    & 44.48 & \best{14.53} & \best{56.92}
    & \best{28.16} & \best{27.86} & \best{24.17}
    & \best{26.01} & \best{64.14} & \best{42.15}
    & \best{36.49} \\
\bottomrule
\end{tabular}%
}
\end{table*}

\paragraph{Complementary strengths.}
The comparison reveals a non-trivial trade-off. \textsc{Qwen3-VL-32B-Instruct} consistently outperforms \textsc{Qwen3-32B} on the visual axis across all three task types, suggesting that the vision-language model carries a stronger internal rendering prior that benefits layout and styling fidelity even on text-only tasks. Conversely, the text-only model retains an advantage on Generation Runnability, indicating more robust code synthesis in scenarios where success depends on clean functional implementation rather than visual grounding.

Overall, the two architectures exhibit complementary strengths: vision-language grounding benefits visual fidelity, while the text-only model can still retain an edge in producing functionally reliable interactive code. This result suggests that stronger multimodal perception does not automatically translate into uniformly better web coding performance, especially when the task is primarily constrained by code reasoning rather than visual reconstruction.

\subsubsection{Impact of Thinking Mode on Performance}
\label{subsec:think_mode}

Recent work on reasoning-enhanced LLMs has introduced ``thinking'' or ``chain-of-thought'' modes that encourage models to reason step-by-step before producing a final answer~\citep{guo2025deepseek}. To investigate whether this paradigm benefits web development tasks, we compare the Instruct and Thinking variants of two Qwen3-VL models that appear in our evaluation: Qwen3-VL-235B-A22B and Qwen3-VL-30B-A3B.

As shown in Table~\ref{tab:main_results}, the impact of thinking mode varies across task types and evaluation dimensions. On \textbf{Generation} tasks, both Thinking variants achieve higher Runnability scores than their Instruct counterparts (63.86 vs.\ 61.26 for 235B; 47.37 vs.\ 41.79 for 30B), indicating that chain-of-thought reasoning helps with code structural correctness. However, the 235B Thinking model suffers a notable Spec Implementation drop (35.02 vs.\ 42.14), while the 30B model shows negligible change (20.87 vs.\ 20.80). On \textbf{Edit} and \textbf{Repair} tasks, differences between Thinking and Instruct variants are relatively minor for both scales.

The limited impact on Edit and Repair tasks likely reflects that these tasks exceed the current capability boundary of Qwen3-VL models. As shown in Table~\ref{tab:main_results}, both models score substantially lower on Edit than on Generation (e.g., 27.74 vs.\ 61.26 on the executability dimension for the 235B Instruct), suggesting that accurately comprehending existing code, locating modification points, and producing precise changes poses a fundamental challenge. When the task difficulty surpasses the model's base competence, thinking mode cannot compensate for the lacking skills—there is insufficient domain knowledge for the reasoning chain to meaningfully build upon.

For Generation tasks, where models perform relatively better, the Spec Implementation degradation of the 235B Thinking variant is notable. Our error analysis reveals that this model produces significantly more \emph{Feature Missing} errors---cases where required interactive features are absent from the output. We attribute this to \textbf{attention dilution} caused by lengthy reasoning chains. Web development prompts often specify multiple requirements simultaneously---layout, styling, interactive behaviors, and responsiveness. The 235B model, with its greater capacity, generates substantially longer thinking chains than its 30B counterpart, pushing the original feature specifications far from the code generation point in the context window. This makes it easier for the model to overlook specific requirements, producing structurally sound but incomplete implementations. The 30B model's shorter reasoning chains preserve proximity to the original prompt, explaining its stable Spec Implementation scores.

\subsubsection{Generation Error Patterns}
\label{subsec:error_analysis}

To understand not only how well but also \emph{how} LLMs fail, we design a structured error analysis framework that classifies every point deduction into a two-level taxonomy spanning four domains (Code Execution, Functional, Visual/Style, and Non-Functional) with fifteen fine-grained error types, and further attributes each error to a root cause. Full taxonomy definitions, the classification prompt, and the decision flowchart are provided in Appendix~\ref{app:error_prompt}; an extended cross-task breakdown is presented in Section~\ref{detailed_error_analysis}.

Across models, three failure modes dominate. \textbf{Feature Missing} is the most common generation error, especially on difficult prompts that combine layout, interaction, and styling constraints. \textbf{Visual inconsistency} remains pervasive even when code executes correctly, confirming the gap between functional correctness and aesthetic fidelity observed in Table~\ref{tab:main_results}. Finally, in repair settings, models often fix the visible symptom while missing the underlying semantic cause, which is consistent with the weak performance on semantic defect categories in the subtask analysis above.

More concretely, Feature Missing and Resource Fail together account for roughly 40\%--55\% of all generation errors across most models. Lower-ranked models accumulate many fundamental failures such as missing functionality and console errors, whereas stronger models reduce these basic failures and leave a larger share of finer-grained layout and styling issues. Distinct modality-specific patterns also emerge: text-conditioned generation is dominated by Feature Missing errors, indicating difficulty translating natural-language specifications into executable interaction logic; image-conditioned generation shifts toward layout, color, and visual-fidelity errors, exposing weakness in pixel-level reproduction; and video-conditioned generation exhibits a more balanced mix of functional and visual errors, reflecting the compound challenge of understanding temporal interaction sequences while faithfully reproducing static appearance. In other words, text primarily stresses requirement comprehension, images stress visual reconstruction, and videos simultaneously stress temporal reasoning and appearance matching, making them the most compositionally demanding input modality.

\begin{figure*}[t]
    \centering
    \includegraphics[width=\linewidth]{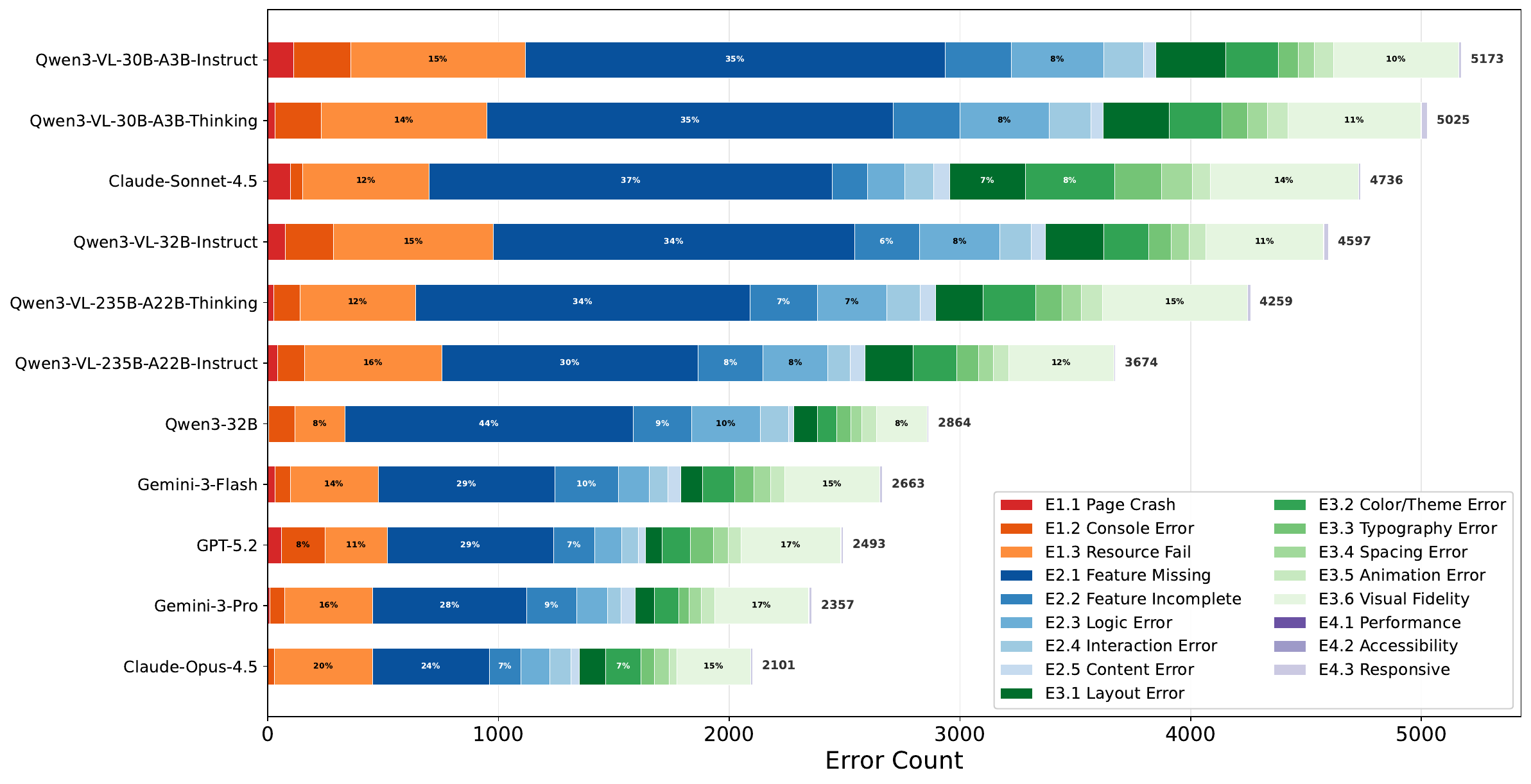}
    \caption{Overall error distribution across all evaluated models on web generation tasks. Feature Missing and Resource Fail dominate across models, while stronger models exhibit a larger fraction of finer-grained visual and styling errors after reducing fundamental execution failures.}
    \label{fig:error_overall}
\end{figure*}

\begin{figure*}[t]
    \centering
    \includegraphics[width=\linewidth]{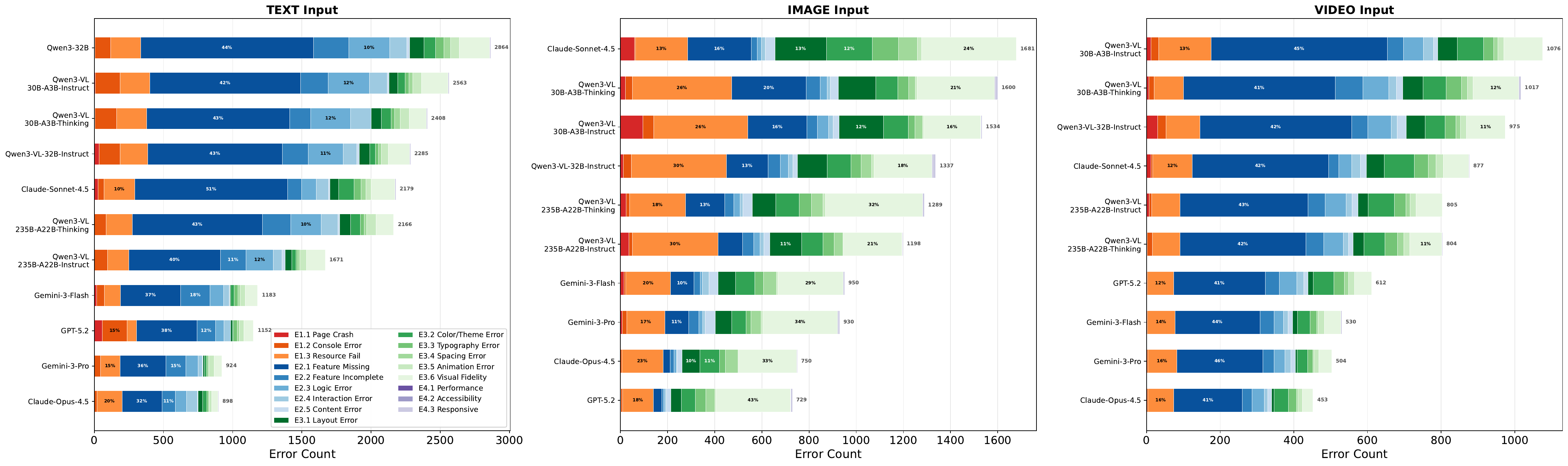}
    \caption{Error distribution by input modality. Text-conditioned generation is dominated by functional omissions, image-conditioned generation shifts toward visual fidelity and layout errors, and video-conditioned generation exhibits a balanced mix of functional and visual failures.}
    \label{fig:error_by_input}
\end{figure*}

\subsubsection{Editing and Repair Error Patterns}
\label{detailed_error_analysis}

The overview above establishes the dominant generation-side failure modes and modality-specific patterns. We next extend the analysis with task-specific quantitative error distributions for Edit and Repair, revealing where patch-based models fail beyond the generation setting.

\begin{figure*}[htbp]
    \centering
    \includegraphics[width=\linewidth]{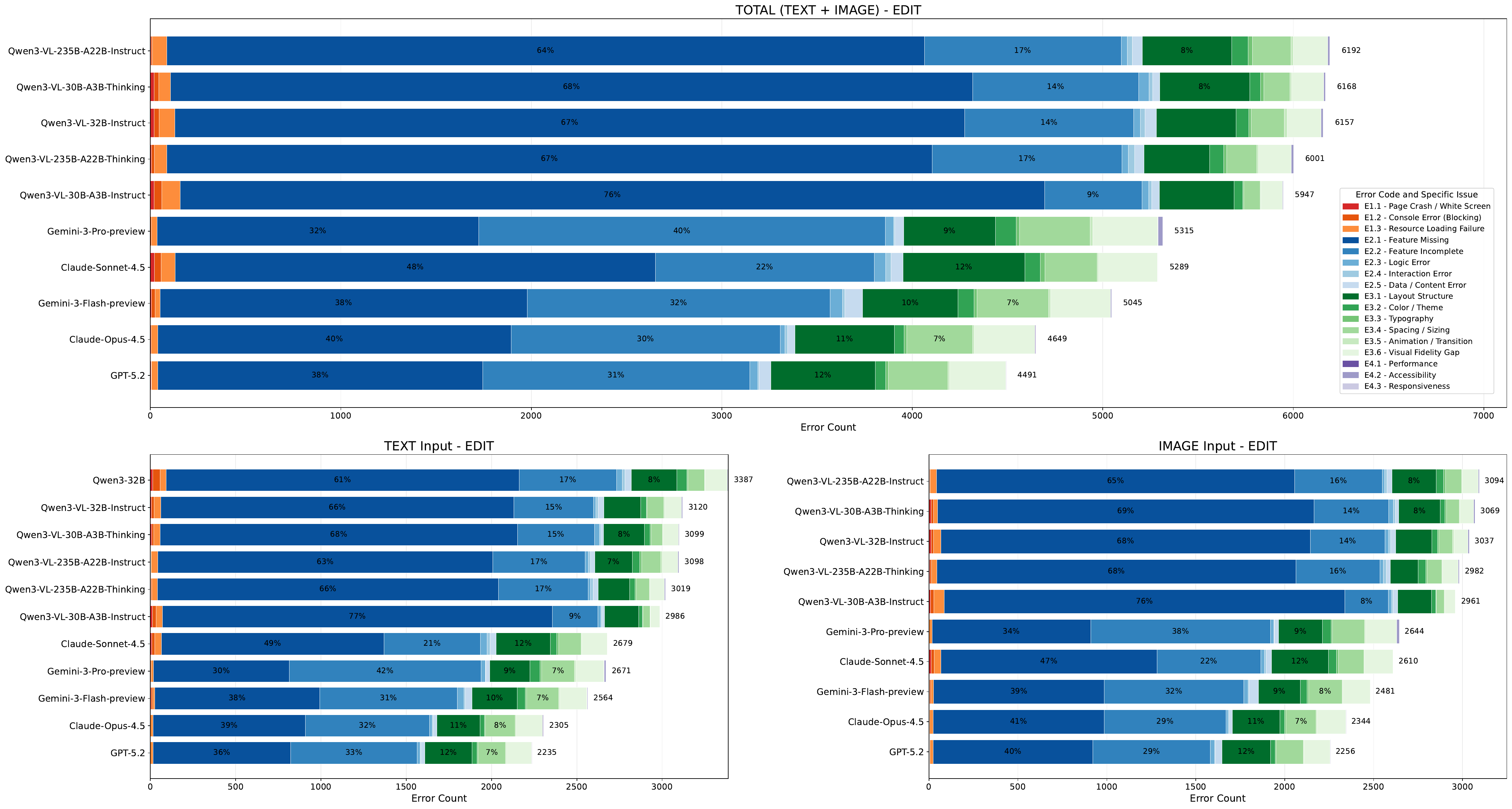}
    \caption{Quantitative distribution of error types for \textbf{Edit tasks}. The errors are categorized into Blocking/Crash (Orange/Red), Logic \& Features (Blue), Visual \& Layout (Green), and Accessibility/Performance (Purple). The total error count for each model is displayed on the right.}
    \label{fig:error_edit}
\end{figure*}

\begin{figure*}[htbp]
    \centering
    \includegraphics[width=\linewidth]{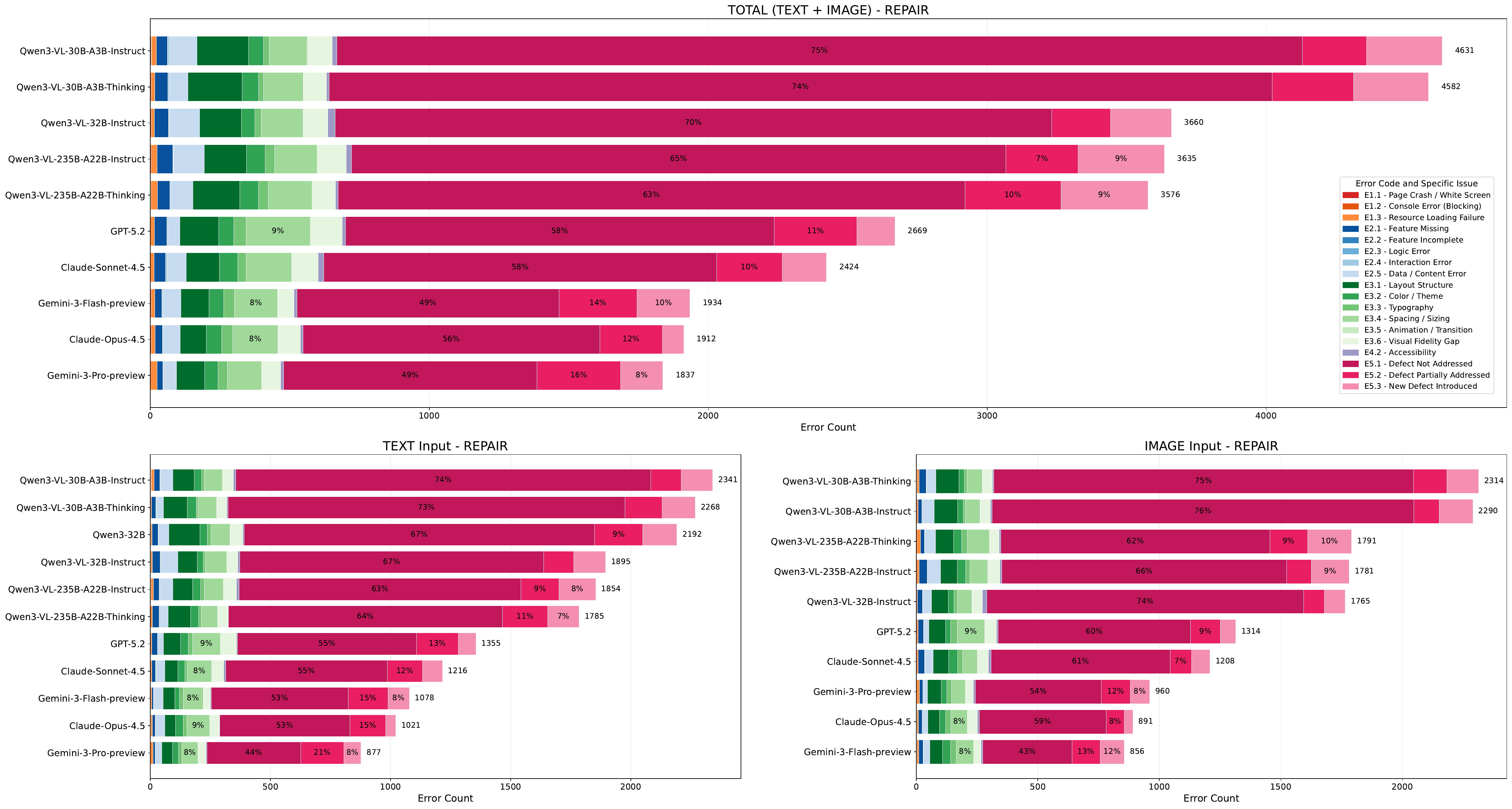}
    \caption{Quantitative distribution of error types for \textbf{Repair tasks}. In addition to standard web errors, Repair tasks introduce defect-resolution specific errors (Pink/Magenta): Defect Not Addressed, Partially Addressed, and New Defect Introduced.}
    \label{fig:error_repair}
\end{figure*}

\paragraph{Quantitative Error Distribution in Edit and Repair Tasks.}
To complement the qualitative observations above, we conduct a comprehensive quantitative breakdown of error frequencies across Edit and Repair tasks. By analyzing the automated checklist deduction logs, we categorize the failures into specific sub-types. Figures~\ref{fig:error_edit} and \ref{fig:error_repair} illustrate the total error counts and their proportional distributions across models for Edit and Repair tasks, respectively.

Several striking patterns emerge from this quantitative lens:

\begin{itemize}[leftmargin=1.5em]
    \item \textbf{Edit Tasks are Bottlenecked by Feature Completeness and Logic (E2).} As shown in Figure~\ref{fig:error_edit}, the vast majority of errors in editing tasks stem from the \textit{Feature Missing (E2.1)} and \textit{Feature Incomplete (E2.2)} categories (represented in dark and medium blue). For open-source models like Qwen3-VL-30B-A3B-Instruct, E2.1 alone accounts for up to 76\% of all checklist failures. Even for top-tier closed-source models, these logical and feature-level omissions dominate (e.g., Claude-Opus-4.5: 40\% E2.1 and 30\% E2.2). This aligns with our qualitative finding that models often suffer from ``partial implementation,'' losing track of complex or multi-step editing instructions.
    
    \item \textbf{Visual Fidelity (E3) is the Secondary Challenge in Editing.} Visual/Layout errors (green segments) form the second-largest block in Edit tasks. Closed-source models notably struggle with \textit{Layout Structure (E3.1)} and \textit{Visual Fidelity Gap (E3.6)}, indicating that while they can write the logical JavaScript, aligning CSS properties precisely with the user's aesthetic intent remains difficult.
    
    \item \textbf{Repair Tasks Fail Primarily due to Unaddressed Defects (E5.1).} Figure~\ref{fig:error_repair} reveals a completely different error paradigm for Repair tasks. The newly introduced repair-specific categories (pink/magenta) overwhelmingly dominate the distribution. Specifically, \textit{Defect Not Addressed (E5.1)} is the primary failure mode. For weaker models (e.g., Qwen3-VL-30B-A3B-Thinking), a staggering 74\% of errors occur because the generated patch simply fails to fix the original bug. Even Gemini-3-Pro-Preview, the absolute best performer in Repair (\S\ref{subsec:main_results}), sees 49\% of its errors coming from E5.1. 
    
    \item \textbf{The ``Over-editing'' Penalty in Repair (E5.3).} We also observe a notable proportion of \textit{New Defect Introduced (E5.3)} errors in Repair tasks (ranging from 8\% to 12\% for closed-source models). This quantitatively corroborates the heavy right-tail distribution observed in our Patch Complexity Analysis (\S\ref{subsec:patch_complexity}): models that generate excessively large patches (like Claude and GPT-5.2) frequently break previously working functionality while attempting to fix a localized bug.
    
    \item \textbf{Modality Consistency.} When splitting the analysis between Text Input and Image Input (bottom panels of both figures), the proportional distribution of error categories remains remarkably stable within each model. This suggests that the core weaknesses—failing to implement complete features in Edit, and failing to locate and fix the defect in Repair—are fundamental reasoning bottlenecks rather than modality-specific perceptual failures.
\end{itemize}
\section{Related Work}
\label{sec:related_work}

Our work is closely related to two research threads: (i) \textbf{code-capable foundation models and agentic coding systems}, and (ii) \textbf{benchmarks and evaluation frameworks for web development} that require judging both visual quality and interactive correctness.

\paragraph{Code LLMs and code agents.}
Large language models for code generation have progressed rapidly from early program synthesis benchmarks~\citep{austin2021program,chen2021evaluating} and competition-level reasoning~\citep{li2022competition} to fully interactive coding agents capable of autonomous software engineering. On the model side, both proprietary systems--- Gemini-3-Pro~\citep{gemini2023gemini}, Claude-Opus-4.5~\citep{claude}---and open-source alternatives--- Qwen3-Coder~\citep{yang2025qwen3} and OpenCoder~\citep{huang2024opencoder}---have demonstrated strong performance on standard code generation benchmarks such as HumanEval~\citep{chen2021evaluating} and LiveCodeBench~\citep{jain2024livecodebench}. On the agent side, SWE-agent~\citep{yang2024sweagent} and OpenHands~\citep{wang2024openhands} equip LLMs with tool-use interfaces for repository-level software engineering, while commercial platforms such as Devin~\citep{cognition2024devin} and Cursor demonstrate the practical viability of agentic coding workflows. SWE-bench~\citep{jimenez2023swe} has become the de facto evaluation framework for these agents, driving rapid progress in automated bug fixing and code editing.
Nevertheless, web development introduces a distinct challenge compared to algorithmic programming or repository-level repair: success is ultimately reflected in the \emph{user-facing artifact}---layout fidelity, design aesthetics, responsiveness, interaction logic, state transitions, and accessibility. These criteria are difficult to capture with purely code-based metrics and can be missed by evaluation suites designed primarily for functional correctness.

\paragraph{Benchmarks for web coding.}
Existing web-coding benchmarks can be categorized along two orthogonal axes: \emph{task type} and \emph{input modality}.

From the task perspective, benchmarks often study:
\begin{itemize}[leftmargin=1.2em]
	\item \textbf{Generation:} producing web pages or mini-apps from requirements. This ranges from early UI-to-code work such as pix2code~\citep{beltramelli2017pix2code} and Web2Code~\citep{yun2024web2code}, to more recent benchmarks including Design2Code~\citep{si2024design2code} for screenshot-to-HTML conversion, WebGen-Bench~\citep{lu2025webgen} for interactive website generation from scratch, DesignBench~\citep{xiao2025designbench} for MLLM-based front-end code generation, and Web-Bench~\citep{xu2025web} for evaluating code against web standards and frameworks. Interaction2Code~\citep{wan2024interaction2code} further extends the modality to interactive prototypes, while IWR-Bench~\citep{chen2025iwr} and FronTalk~\citep{wu2025frontalk} explore video-conditioned and conversational generation settings, respectively.
	\item \textbf{Editing:} modifying an existing codebase to satisfy new requirements. SWE-bench Multimodal~\citep{yang2024swebenchmultimodal} extends the original SWE-bench~\citep{jimenez2023swe} to visual software domains, requiring models to interpret screenshots alongside issue descriptions.
	\item \textbf{Repair:} fixing defects in UI/UX or broken interactions, ranging from text-described bugs to visually grounded defect descriptions.
\end{itemize}
Several recent efforts aim at holistic, multi-dimensional coverage. WebUIBench~\citep{lin2025webuibench} benchmarks WebUI-to-code generation with comprehensive metrics; FullFront~\citep{sun2025fullfront} spans the full front-end engineering workflow; ArtifactsBench~\citep{zhang2025artifactsbench} bridges the visual-interactive gap in LLM code generation evaluation; and WebDev Arena~\citep{lmsys2024webdev} provides a human-preference-based leaderboard for web development. WebCoderBench~\citep{liu2026webcoderbench} proposes comprehensive and interpretable evaluation metrics, while WebMMU~\citep{awal2025webmmu} extends coverage to multilingual website understanding.

Despite this growing body of work, most existing benchmarks focus on a single task type (typically generation) or a single input modality (typically text or static images), and their evaluations often rely on either weak proxies (e.g., single screenshot similarity or DOM heuristics) or brittle scripted tests that require strict attribute conventions. WebCompass addresses these limitations by spanning three modalities, three task types, and employing execution-grounded evaluation that tests end-to-end runtime behavior.

\paragraph{Evaluation paradigms for interactive visual artifacts.}
Evaluation methods for web-facing artifacts commonly fall into three classes:
\begin{itemize}[leftmargin=1.2em]
	\item \textbf{Rule-/test-based evaluation.} Deterministic test suites, as employed by SWE-bench~\citep{jimenez2023swe} and Web-Bench~\citep{xu2025web}, provide precise and reproducible verdicts but typically require heavy instrumentation, strict naming conventions, and substantial engineering effort to achieve good coverage across diverse implementations.
	\item \textbf{Agent-based interaction.} Web agents---as pioneered by WebArena~\citep{zhou2023webarena} and extended to multimodal settings in VisualWebArena~\citep{koh2024visualwebarena}---can explore an artifact by interacting with the page and checking outcomes. However, coverage remains challenging: predefined action spaces may miss complex behaviors, and long-horizon workflows are hard to validate end to end.
	\item \textbf{LLM/MLLM-as-a-Judge.} Language or multimodal referees~\citep{zheng2023judging,ge2023mllm} can scale to open-ended designs and assess multiple dimensions jointly, but may be subjective without careful rubric design and evidence grounding.
\end{itemize}

Our work adopts a task-aware combination of these paradigms. For \textbf{editing} and \textbf{repair}, we use \emph{checklist-guided} LLM-as-a-Judge~\citep{zheng2023judging} to anchor evaluation in per-task, fine-grained criteria with evidence grounding. For \textbf{generation}, where acceptable solutions are diverse and interactivity is open-ended, we introduce an \emph{Agent-as-a-Judge} protocol~\citep{zhuge2024agent} that combines browser-based interaction (via the Model Context Protocol) with iterative test-case synthesis, providing stronger and more realistic validation than any single evaluation paradigm alone.

\section{Conclusion}
\label{sec:conclusion}

We presented \textbf{WebCompass}, a multimodal benchmark unifying generation, editing, and repair across text, image, and video modalities, with task-aware evaluation combining LLM-as-a-Judge and Agent-as-a-Judge protocols to assess executability, functional behavior, and visual quality across task-specific rubrics. Our experiments reveal that closed-source models lead by $\sim$25 points over the best open-source alternatives, visual quality remains the most persistent bottleneck even for frontier models, and generation, editing, and repair stress fundamentally different capabilities---no single model dominates all three. These findings suggest that advancing web coding agents requires not only stronger functional reasoning but also deeper visual design understanding and greater output consistency, pointing toward a future where coding agents are evaluated---and optimized---as holistic builders of user-facing experiences rather than mere code generators.

\bibliographystyle{unsrtnat}
\bibliography{ArtifactsBench} 

\appendix
\section{Appendix}

\subsection{Limitations}
\label{subsec:limitations}

While WebCompass represents a substantial step toward comprehensive evaluation of web coding agents, we acknowledge several limitations that future work should address.

\paragraph{Front-end focus.}
WebCompass concentrates exclusively on front-end web development (HTML, CSS, JavaScript, and front-end frameworks). It does not evaluate back-end capabilities such as database design, server-side logic, API development, or deployment workflows. Real-world web engineering involves full-stack development, and extending the benchmark to cover back-end tasks would provide a more complete assessment.

\paragraph{Structured queries vs.\ creative intent.}
For generation tasks, we deliberately refine underspecified user queries into structured web design documents (specifying content, interaction, and visual appearance) to enable reproducible, automated evaluation. This design choice means that our benchmark primarily tests \emph{instruction-following} capability rather than the ability to interpret vague, creative intent. We acknowledge this as an inherent trade-off: WebCompass prioritizes \emph{deterministic evaluation standards over open-ended creative assessment}. Complementary benchmarks that explicitly measure creative divergence (e.g., via human preference ranking) would provide a valuable additional perspective.

\paragraph{Limited real-time interaction with dynamic web pages.} Our evaluation protocols currently cannot perform real-time interaction with highly dynamic web pages---such as browser-based games or applications with frequent state transitions---in the way a human would. While our framework supports both natural interactions and script-based inspection, it remains challenging to keep pace with rapidly evolving page states, making it difficult to accurately assess time-sensitive behaviors such as real-time game logic, continuous animation responses, or state transitions that depend on precise timing. As a result, the evaluation quality for such highly dynamic web pages may not fully reflect their actual functionality and user experience.

\paragraph{Static benchmark and contamination risk.}
As a static benchmark, WebCompass is susceptible to data contamination if future models are trained on data that includes our tasks or similar web pages. While we mitigate this through diverse data sources and original task synthesis, maintaining a contamination-free evaluation over time may require periodic updates or dynamic task generation.

\paragraph{Evaluation cost.}
The Agent-as-a-Judge protocol, while more thorough than static evaluation, is computationally expensive. Each generation task requires launching a headless browser, executing multi-step interaction sequences, and synthesizing iterative test cases, which significantly increases evaluation time and cost compared to simpler metrics. This may limit the benchmark's accessibility for resource-constrained research groups.

\subsection{Disclosure of LLM Assistance}

The authors independently conceived and executed all scientific ideas, algorithmic implementations, and experimental data analyses. Large Language Models (LLMs) were employed exclusively as auxiliary tools for language editing and enhancing the clarity of the manuscript. No experimental data, training samples, or reported results were generated using LLMs without rigorous human verification.

% \subsection{Experimental Hyperparameters}
% \label{app:hyperparameters}

% Temperature and top-$p$ are tuned according to each model's official recommendations; the maximum output token limit is set sufficiently large to prevent truncation.

\subsection{Per-Dimension Framework Evaluation}
\label{app:framework_detail}

Table~\ref{tab:framework_subset} provides the per-dimension breakdown of the framework subset evaluation summarized in Section~\ref{subsec:framework_subset}. Dimension abbreviations follow Table~\ref{tab:main_results}.

\begin{table*}[h]
\centering
\caption{Per-dimension subset evaluation across different front-end frameworks. Each model is tested on 60 randomly sampled tasks per category using React, Vue, and Vanilla HTML/JS. \best{Green bold}: best; \second{blue underline}: second best per framework.}
\label{tab:framework_subset}
\resizebox{\textwidth}{!}{%
\renewcommand{\arraystretch}{1.1}
\small
\begin{tabular}{@{}l|l|ccc|ccc|ccc@{}}
\toprule
\multirow{2}{*}{\textbf{\textcolor{headerblue}{Model}}}
 & \multirow{2}{*}{\textbf{\textcolor{headerblue}{FW}}}
 & \multicolumn{3}{c|}{\textbf{\textcolor{headerblue}{Generation}}}
 & \multicolumn{3}{c|}{\textbf{\textcolor{headerblue}{Edit}}}
 & \multicolumn{3}{c}{\textbf{\textcolor{headerblue}{Repair}}} \\
\cmidrule(l){3-11}
 & & \textbf{\textcolor{headerblue}{RUN.}}
   & \textbf{\textcolor{headerblue}{SPI.}}
   & \textbf{\textcolor{headerblue}{DSQ.}}
   & \textbf{\textcolor{headerblue}{ITG.}}
   & \textbf{\textcolor{headerblue}{FTI.}}
   & \textbf{\textcolor{headerblue}{STC.}}
   & \textbf{\textcolor{headerblue}{RCT.}}
   & \textbf{\textcolor{headerblue}{ITI.}}
   & \textbf{\textcolor{headerblue}{RFF.}} \\
\midrule
\multirow{3}{*}{\rotatebox{0}{\textbf{GPT-5.2}}}
 & React   & 62.08 & \best{60.57} & \second{47.88} & 43.60 & 40.10 & 35.86 & \best{54.58} & \best{82.40} & \best{62.56} \\
 & Vue     & \second{65.08} & \second{56.79} & 45.29 & \second{45.73} & \second{43.30} & \second{38.38} & \second{48.85} & 76.77 & 59.14 \\
 & Vanilla & \best{75.13} & 60.18 & \best{56.38} & \best{65.20} & \best{61.87} & \best{55.79} & 44.10 & \second{79.82} & \second{60.34} \\
\midrule
\multirow{3}{*}{\rotatebox{0}{\textbf{Gemini-3-Pro-Preview}}}
 & React   & 61.05 & 47.29 & 46.11 & \second{54.37} & \second{50.31} & \second{43.92} & \second{44.02} & \second{81.67} & \second{64.12} \\
 & Vue     & \second{71.01} & \best{55.98} & \second{59.61} & 43.58 & 39.13 & 34.23 & 39.25 & 74.70 & 57.11 \\
 & Vanilla & \best{75.02} & \second{55.70} & \best{64.49} & \best{74.84} & \best{69.48} & \best{62.06} & \best{50.95} & \best{86.62} & \best{65.78} \\
\midrule
\multirow{3}{*}{\rotatebox{0}{\textbf{Claude-Opus-4.5}}}
 & React   & \second{79.16} & \best{71.91} & 56.78 & \second{55.84} & \second{49.10} & \second{42.24} & \second{44.47} & 78.37 & \second{63.97} \\
 & Vue     & 72.85 & 66.99 & \second{58.74} & 40.14 & 38.27 & 30.04 & 41.14 & \second{81.94} & 62.85 \\
 & Vanilla & \best{79.23} & \second{71.72} & \best{62.59} & \best{78.31} & \best{72.47} & \best{65.37} & \best{47.21} & \best{85.56} & \best{64.38} \\
\midrule
\multirow{3}{*}{\rotatebox{0}{\textbf{Qwen3-VL-235B-A22B-Instruct}}}
 & React   & \second{45.76} & \second{25.80} & \second{36.11} & \second{20.57} & \second{18.15} & \second{15.84} & \best{26.16} & \second{62.63} & \second{45.33} \\
 & Vue     & 42.18 & 23.17 & 27.46 & 15.12 & 16.02 & 13.31 & 22.22 & 57.00 & 37.56 \\
 & Vanilla & \best{62.16} & \best{39.97} & \best{45.01} & \best{26.33} & \best{24.99} & \best{22.19} & \second{25.09} & \best{71.46} & \best{46.89} \\
\bottomrule
\end{tabular}%
}
\end{table*}

\subsection{Model Card}
\label{model_card}
Table~\ref{tab:model_list} lists all model variants referenced in the experiments, including the auxiliary comparison model used in Section~\ref{subsec:text_vs_vlm}.
\begin{table}[htbp]
\centering
\caption{List of model variants referenced in the experiments.}
\label{tab:model_list}
\begin{tabular}{@{}l@{}}
\toprule
\textbf{\textcolor{headerblue}{Model}} \\
\midrule
Claude-Opus-4.5~\citep{claude} \\
Claude-Sonnet-4.5~\citep{claude} \\
Gemini-3-Pro-Preview~\citep{gemini} \\
Gemini-3-Flash-Preview~\citep{gemini} \\
GPT-5.2~\citep{openai} \\
Qwen3-32B~\citep{yang2025qwen3} \\
Qwen3-VL-32B-Instruct~\citep{bai2025qwen3} \\
Qwen3-VL-235B-A22B-Instruct~\citep{bai2025qwen3} \\
Qwen3-VL-235B-A22B-Thinking~\citep{bai2025qwen3} \\
Qwen3-VL-30B-A3B-Instruct~\citep{bai2025qwen3} \\
Qwen3-VL-30B-A3B-Thinking~\citep{bai2025qwen3} \\
\bottomrule
\end{tabular}%
\end{table}

\subsection{Detailed Worst-of-$n$ Stability Results}

\label{app:worst_of_n}

Table~\ref{tab:worst_of_n} extends the Worst-of-$n$ stability analysis presented in the main text by reporting $Pass@1$, $W@2$, and $W@4$ scores across all nine evaluation dimensions, grouped by task category.

\begin{table*}[htbp]
\centering
\caption{Consistency and Stability Results for Gemini-3-Pro-Preview and Qwen3-VL-235B-A22B-Instruct ($n=4$). $Pass@1$, $W@2$, and $W@4$ are reported across all nine evaluation dimensions grouped by task category; $\Delta\!\downarrow$ denotes the relative drop from $Pass@1$ to $W@4$.}
\label{tab:worst_of_n}
\resizebox{\textwidth}{!}{%
\renewcommand{\arraystretch}{1.25}
\begin{tabular}{@{}l|ccc|ccc|ccc@{}}
\toprule
\multirow{2}{*}{\textbf{\textcolor{headerblue}{Metric}}}
 & \multicolumn{3}{c|}{\textbf{\textcolor{headerblue}{Generation}}}
 & \multicolumn{3}{c|}{\textbf{\textcolor{headerblue}{Editing}}}
 & \multicolumn{3}{c}{\textbf{\textcolor{headerblue}{Repair}}} \\
\cmidrule(lr){2-4}\cmidrule(lr){5-7}\cmidrule(l){8-10}
 & \textbf{\textcolor{headerblue}{RUN.}}
 & \textbf{\textcolor{headerblue}{SPI.}}
 & \textbf{\textcolor{headerblue}{DSQ.}}
 & \textbf{\textcolor{headerblue}{ITG.}}
 & \textbf{\textcolor{headerblue}{FTI.}}
 & \textbf{\textcolor{headerblue}{STC.}}
 & \textbf{\textcolor{headerblue}{RCT.}}
 & \textbf{\textcolor{headerblue}{ITI.}}
 & \textbf{\textcolor{headerblue}{RFF.}} \\
\midrule
% ---- Gemini-3-Pro ----
\rowcolor{closedtag}
\multicolumn{10}{c}{\textit{\textcolor{headerblue}{\textbf{Gemini-3-Pro-Preview}}}} \\
\midrule
\rowcolor{white} Pass@1 (\%)          & 75.31 & 56.90 & 60.68 & 74.84 & 69.48 & 62.06 & 50.95 & 86.62 & 65.78 \\
\rowcolor{rowgray} W@2 (\%)           & 73.55 & 47.74 & 56.03 & 67.41 & 62.41 & 53.77 & 39.83 & 80.41 & 58.19 \\
\rowcolor{white} W@4 (\%)             & 68.92 & 39.80 & 49.27 & 63.45 & 57.89 & 49.51 & 31.31 & 72.07 & 49.80 \\
\rowcolor{rowgray} $\Delta\!\downarrow$ (\%) & 8.48 & 30.05 & 18.80 & 15.22 & 16.68 & 20.22 & 38.55 & 16.80 & 24.29 \\
\midrule
% ---- Qwen3-VL-235B-Instruct ----
\rowcolor{closedtag}
\multicolumn{10}{c}{\textit{\textcolor{headerblue}{\textbf{Qwen3-VL-235B-A22B-Instruct}}}} \\
\midrule
\rowcolor{white} Pass@1 (\%)          & 61.22 & 38.59 & 42.77 & 26.33 & 24.99 & 22.19 & 25.09 & 71.46 & 46.89 \\
\rowcolor{rowgray} W@2 (\%)           & 52.75 & 28.42 & 34.57 & 19.54 & 17.62 & 17.64 & 21.00 & 62.46 & 38.68 \\
\rowcolor{white} W@4 (\%)             & 44.42 & 23.56 & 29.09 & 15.91 & 14.50 & 13.76 & 17.66 & 58.23 & 32.86 \\
\rowcolor{rowgray} $\Delta\!\downarrow$ (\%) & 27.44 & 38.95 & 31.99 & 39.57 & 41.98 & 37.99 & 29.61 & 18.51 & 29.92 \\
\bottomrule
\end{tabular}%
}
\end{table*}

\subsection{Prompt Templates}
\label{app:prompts}

This section presents the prompt templates used in our pipeline, covering task prompts for code generation, editing, and repair (\S\ref{subsec:generation_prompts}--\ref{subsec:repair_prompts}), evaluation prompts for LLM-as-a-Judge and Agent-as-a-Judge (\S\ref{subsec:judge_prompts}--\ref{subsec:agent_judge_prompt}), and auxiliary prompts for checklist generation and error analysis (\S\ref{subsec:checklist_prompt}--\ref{app:error_prompt}).

\subsubsection{Error Analysis Prompt}
\label{app:error_prompt}

Below is the prompt used to classify point deductions into standardized error types and root causes. The full prompt includes the complete taxonomy definitions, a decision flowchart, point allocation rules, and worked examples.

\begin{promptbox}{Error Analysis Prompt (Part 1: Taxonomy \& Rules)}
You are a web development QA analyst. Classify each point deduction
from LLM-generated web pages into standardized error types.

== Error Type Taxonomy (3-level) ==

E1 - Code Execution Errors
  E1.1 | Page Crash / White Screen | Page fails to render entirely
  E1.2 | Console Error (Blocking)  | JS error that prevents feature from working
  E1.3 | Resource Loading Failure  | Missing images, fonts, external dependencies

E2 - Functional Errors
  E2.1 | Feature Missing     | A required feature/component is entirely absent
  E2.2 | Feature Incomplete  | Feature exists but is only partially implemented
  E2.3 | Logic Error         | Feature exists but produces wrong behavior/output
  E2.4 | Interaction Error   | Wrong event type, trigger condition, or user flow
  E2.5 | Data/Content Error  | Wrong text, numbers, items, or dynamic content

E3 - Visual/Style Errors
  E3.1 | Layout Structure Error      | Wrong arrangement, grid, flex direction, positioning
  E3.2 | Color/Theme Error           | Wrong colors, gradients, or visual theme
  E3.3 | Typography Error            | Wrong font, size, weight, line-height
  E3.4 | Spacing/Sizing Error        | Wrong margin, padding, width, height
  E3.5 | Animation/Transition Error  | Missing, wrong, or poorly executed animations
  E3.6 | Visual Fidelity Gap         | General mismatch with reference image/video

E4 - Non-Functional Errors
  E4.1 | Performance Issue      | Lag, high CPU, memory leak, slow rendering
  E4.2 | Accessibility Issue    | Missing alt text, poor contrast, no keyboard nav
  E4.3 | Responsiveness Issue   | Breaks at different viewport sizes

== Root Cause Analysis (select the MOST specific one) ==
  Requirement Misunderstanding | Prompt clearly stated X, LLM built Y
  Requirement Omission        | LLM skipped/forgot an explicit requirement
  Insufficient Reproduction   | Visual/behavioral gap vs reference image/video
  Capability Limitation       | LLM lacks skill for this technique
  Hallucination               | LLM used non-existent APIs or fabricated behavior
  Oversimplification          | LLM took shortcuts, used simpler approach
  null                        | Cannot determine or none of the above

== Decision Flowchart ==
For each deducted point:
1. Does the page crash or show blocking console errors? -> E1.x
2. Does a required feature not work as specified?       -> E2.x
3. Does it work but look wrong?                         -> E3.x
4. Does it look right but have non-functional issues?   -> E4.x

== Points Allocation Rule ==
When a single checklist item has multiple issues mentioned in reason:
  - Allocate points proportionally to issue severity
  - If unclear, split evenly among identified issues
  - Critical failures get more points than minor issues

== Output Format ==
Return a JSON array:
[{"checklist_id": <id>, "task": "<task>",
  "score": <score>, "max_score": <max_score>,
  "errors": [{"type": "<E1.1|...|E4.3>",
    "description": "<concise description>",
    "points_deducted": <number>,
    "root_cause": "<label|null>"}]}]

== Rules ==
1. sum(points_deducted) must equal max_score - score for each item
2. Each error gets exactly ONE type code
3. Full-score items -> empty errors array
4. Multiple distinct issues -> separate error objects
\end{promptbox}

\begin{promptbox}{Error Analysis Prompt (Part 2: Few-Shot Examples)}
== Example 1: Runtime error ==
Input: {"id":1, "task":"Page loads correctly",
  "max_score":5, "score":0,
  "reason":"Uncaught ReferenceError: initApp is not defined.
  Page shows white screen."}
Output:
[{"checklist_id":1, "task":"Page loads correctly", "score":0,
  "max_score":5, "errors":[{"type":"E1.2",
  "description":"Uncaught ReferenceError prevents page init",
  "points_deducted":5, "root_cause":"Hallucination"}]}]

== Example 2: Multiple issues in one item ==
Input: {"id":3, "task":"Test mouse wake effect",
  "max_score":12, "score":6,
  "reason":"Wake works on mousedown+drag but NOT on
  hover-only movement. Dissipation too slow (visible after 3s)."}
Output:
[{"checklist_id":3, "task":"Test mouse wake effect", "score":6,
  "max_score":12, "errors":[
    {"type":"E2.4",
     "description":"Wake requires mousedown instead of hover",
     "points_deducted":4,
     "root_cause":"Requirement Misunderstanding"},
    {"type":"E2.3",
     "description":"Wake dissipation too slow (>3s)",
     "points_deducted":2, "root_cause":null}]}]

== Example 3: Style issue ==
Input: {"id":5, "task":"Check header layout",
  "max_score":6, "score":4,
  "reason":"Logo and nav are present but laid out vertically
  instead of horizontal row."}
Output:
[{"checklist_id":5, "task":"Check header layout", "score":4,
  "max_score":6, "errors":[{"type":"E3.1",
  "description":"Header elements stacked vertically
  instead of required horizontal layout",
  "points_deducted":2,
  "root_cause":"Insufficient Reproduction"}]}]

Now analyze the following checklist:
[CHECKLIST]
\end{promptbox}

\subsubsection{Checklist Generation Prompt}
\label{subsec:checklist_prompt}

Below is the prompt used to generate evaluation checklists. The prompt instructs an LLM to produce structured checklist items spanning three dimensions: Runnability, Spec Implementation, and Design Quality. We show the core instructions and an abridged example; the full prompt includes a complete 13-item few-shot example.

\begin{promptbox}{Checklist Generation Prompt (Part 1: Role \& Format)}
You are a senior and extremely detail-oriented code review expert.
You are proficient in multiple programming languages, frontend
technologies, interaction design, and UI aesthetics. Your task is
to generate a checklist for evaluating responses to the given
[query].

## Role Definition
- Responsibility: Act as a member of an authoritative technical
  review committee--objective, comprehensive, and impartial.
- Attitude: Meticulous, professional, and uncompromising.
- Aesthetic Standards: Excellent design taste and high standards
  for user experience.

## Output Format Requirements (Each checklist item must include)
- task: A clear, single-purpose task for the UI agent to verify.
- category: One of: Runnability | Spec Implementation | Design Quality
- operation_sequence: Steps the UI agent should perform (2-4
  steps, verifiable through screenshots/screen recordings).
- expected_result: Specific, observable success criteria.
- criteria: Strict scoring rules (clear pass/fail thresholds).
- max_score: Maximum score for this item (must be a number).
\end{promptbox}

\begin{promptbox}{Checklist Generation Prompt (Part 2: Dimensions)}
## Evaluation Dimensions and Checklist Structure

### 1. Runnability (Fixed 1 item, worth 10 points)
Fixed universal checklist item:
{"task": "Does the page load correctly and run without errors?",
 "category": "Runnability",
 "operation_sequence": "1. Open Console panel 2. Load the page
   3. Check for red error messages 4. Check Network for 404/500",
 "expected_result": "Page loads completely, no Console errors,
   no failed requests, all static resources load successfully",
 "criteria": "Full 10 pts; JS errors -5; Resource 404 -3;
   White screen = 0; Warnings do not deduct",
 "max_score": 10}

### 2. Spec Implementation (6-10 items, worth 60-70 points)
Core part -- generated entirely based on the Query:
- First 4-5 items: core functionality, main interactions,
  key user journeys (most stringent, hardest to pass)
- Remaining items: secondary features, edge cases, robustness
- Score per item: 8-15 points, allocated by importance

### 3. Design Quality (1 universal + 1-2 Query-specific, 20-25 pts)
Universal item evaluates: color harmony, layout spacing (8px
multiple principle), typography (body font >=14px, line-height
>1.5x). Query-specific items check domain visuals (e.g., game
board design, chart readability, animation smoothness).

## Important Notes
- Hard requirement: sum of all max_score must equal 100
- Hard requirement: 10-16 checklist items total
- Each item must be verifiable through screenshots/console/
  network panel/visual alignment checks
\end{promptbox}

\begin{promptbox}{Checklist Generation Prompt (Part 3: Abridged Example)}
## Complete Example (abridged -- showing 3 of 13 items)
Query: [Multiplayer online chess game with lobby, game board,
  result modal, and profile pages...]

[{"task": "Does the page load correctly and run without errors?",
  "category": "Runnability",
  "operation_sequence": "1. Open Console 2. Load page ...",
  "expected_result": "Page loads, no errors, all resources OK",
  "criteria": "Full 10; JS errors -5; 404 -3; crash = 0",
  "max_score": 10},
 {"task": "Test chessboard piece selection and valid move
   indicators functionality",
  "category": "Spec Implementation",
  "operation_sequence": "1. Start game 2. Click white pawn (e2)
   3. Verify highlights and move indicators 4. Click valid
   square to confirm move with animation",
  "expected_result": "Piece highlighted, dots on empty squares,
   rings on captures, smooth ~200ms animation",
  "criteria": "Full 12; no highlight -4; no indicators -4;
   no animation -2; wrong move -2",
  "max_score": 12},
 {"task": "Verify dark theme with correct primary/accent colors",
  "category": "Design Quality",
  "operation_sequence": "1. Screenshot Lobby and Game pages
   2. Verify dark gradient background 3. Check primary/accent
   colors 4. Verify semi-transparent cards",
  "expected_result": "Dark gradient bg, wood-brown primary,
   royal blue interactive, gold accents, cohesive dark theme",
  "criteria": "Full 5; light theme -5; inconsistent colors -2",
  "max_score": 5},
  ... (10 more items omitted for brevity)]

Query:
---
[QUERY]
---
\end{promptbox}

\subsubsection{Generation Prompts}
\label{subsec:generation_prompts}

We use three generation prompts corresponding to the three input modalities: text, image, and video. All share a common output contract requiring pure Markdown with fenced code blocks. Below we present each variant.

\begin{promptbox}{Text-Guided Generation Prompt}
You are a highly skilled professional front-end engineer.

Your task: Based on the web design document below, generate a
complete runnable web project repository.

Hard output contract (MUST follow):
1) Your entire response MUST be pure Markdown text.
2) ABSOLUTELY NO explanations, no extra commentary.
3) Every file MUST be emitted using the following format:
     # path/to/file.ext
     ```ext
     <full file content>
     ```
4) The heading line MUST start with '# ' followed by the
   file path (relative path).
5) The code fence language MUST match the file type.
6) Include all necessary files so the project can run.
7) Do NOT nest triple backticks inside code blocks.

Few-shot examples:

  # index.html
  ```html
  <!doctype html>
  <html>
    <head>
      <meta charset="utf-8" />
      <title>Demo</title>
      <link rel="stylesheet" href="styles.css" />
    </head>
    <body>
      Hello
      <script type="module" src="main.js"></script>
    </body>
  </html>
  ```

  # styles.css
  ```css
  body { font-family: system-ui; }
  ```

  # main.js
  ```js
  console.log('ok')
  ```

Web design document:
---
[DOCUMENT]
---
\end{promptbox}

\begin{promptbox}{Vision-Guided Generation Prompt}
You are a highly skilled professional front-end engineer.

Your task: Based on the web design document and the reference
screenshots provided, generate a complete runnable web project
repository. The screenshots represent the target UI and function.
The generated site should match the screenshots as closely as
possible.

Hard output contract (MUST follow):
1-7) [Same output contract as Text-Guided Generation]

Startup requirements:
- MUST include a README.md with the simplest way to run locally.
- MUST be runnable via a static server (no backend).
  Prefer Vite or plain static files.

Web design document:
---
[DOCUMENT]
---
\end{promptbox}

\begin{promptbox}{Video-Guided Generation Prompt (Part 1: Analysis Protocol)}
You are a world-class front-end engineer with expertise in
creating pixel-perfect web reproductions. Your task is to analyze
these video frames and create a complete, production-ready web
project repository that exactly replicates the demonstrated
interface with professional-grade quality.

MISSION: Create a flawless HTML reproduction that matches the
video demonstration in every detail, achieving 95%+ quality score.

COMPREHENSIVE ANALYSIS PROTOCOL:

1. TEMPORAL SEQUENCE ANALYSIS:
   - Study frame progression to understand user interactions
   - Identify animation sequences, timing, and easing patterns
   - Map state transitions and user feedback mechanisms
   - Recognize loading states, hover effects, micro-interactions
   - Document exact timing and duration of animations

2. VISUAL DESIGN EXTRACTION:
   - Extract precise color values (prefer hex codes: #RRGGBB)
   - Identify typography: families, sizes, weights, line heights
   - Measure spacing: margins, padding, gaps (use rem/em units)
   - Analyze shadows: box-shadow values, blur, spread, inset
   - Document border radius, opacity, and gradient effects
   - Note z-index layering and stacking contexts

3. LAYOUT & STRUCTURE ANALYSIS:
   - Identify layout systems: Flexbox, CSS Grid, or positioning
   - Map responsive breakpoints and mobile adaptations
   - Document component hierarchy and nesting structure
   - Analyze alignment, distribution, and spacing patterns

4. INTERACTION PATTERN RECOGNITION:
   - Button states: normal, hover, active, focus, disabled
   - Animation triggers: click, hover, scroll, load events
   - State management: data flow and component updates
   - User feedback: visual confirmations and error states
\end{promptbox}

\begin{promptbox}{Video-Guided Generation Prompt (Part 2: Implementation)}
TECHNICAL IMPLEMENTATION REQUIREMENTS:

HTML5 STRUCTURE (MANDATORY):
  - Use semantic HTML5 elements (main, section, article, nav)
  - Include proper meta tags: charset, viewport, description
  - Implement accessibility: ARIA labels, alt text, roles
  - Ensure valid HTML5 markup

CSS IMPLEMENTATION (MANDATORY):
  - Use CSS custom properties (--variables) for theming
  - Implement modern layout: Flexbox and CSS Grid
  - Create smooth animations with cubic-bezier easing
  - Use transform3d for hardware acceleration
  - Implement responsive design with mobile-first approach
  - Include CSS reset/normalize for consistency

JAVASCRIPT FUNCTIONALITY (MANDATORY):
  - Write modern ES6+ with proper error handling (try/catch)
  - Use requestAnimationFrame for smooth animations
  - Add performance optimizations: debouncing, throttling
  - Implement proper event delegation and cleanup

ANIMATION & INTERACTION STANDARDS (MANDATORY):
  - Use CSS transforms for performance (translate3d, scale)
  - Implement 60fps smooth animations
  - Add appropriate transition durations (200-500ms)
  - Create natural easing: ease-out entrance, ease-in exit

OUTPUT SPECIFICATIONS (CRITICAL):
  - Your entire response MUST be pure Markdown text.
  - ABSOLUTELY NO explanations or extra commentary.
  - Every file MUST use: # path/to/file.ext  ```ext ... ```
  - Keep assets inline; avoid external dependencies.

Begin your comprehensive analysis and create the ultimate
reproduction that achieves 95%+ quality score.
\end{promptbox}

\subsubsection{Editing Prompts}
\label{subsec:editing_prompts}

The text-guided and vision-guided editing tasks share the same system prompt. The only difference is that the vision-guided variant additionally includes current-state screenshots in the user message. We present the shared system prompt once, followed by the two user-message variants.

\begin{promptbox}{Editing System Prompt (shared by Text \& Vision variants)}
You are an expert frontend developer. Your task is to edit the
provided web code based on the given instructions.
You will receive the current code and a set of editing instructions.

**Output Format Requirements:**
- Use search/replace blocks to indicate modifications
- Each block must be wrapped in <search_replace path="...">
  </search_replace> tags
- The `path` attribute must specify the relative file path
  (e.g., "index.html", "resources/style.css")
- Each block must contain one <search> and one <replace>

Return XML format with the following structure:
<search_replace path="path/to/file">
<search>
exact text to find in the original file
</search>
<replace>
replacement text with the modification applied
</replace>
</search_replace>

If you want to create additional files:
<search_replace path="path/to/new_file">
<search></search>
<replace>
complete code for the new file
</replace>
</search_replace>
The search block for new files should be empty.

Important:
- The <search> block must contain the EXACT text from the
  original file (including whitespace and indentation).
- The <replace> block contains the modified code.
- One <search_replace> block can only contain one pair of
  <search> and <replace>.
- You can include multiple <search_replace> blocks if you
  need to modify multiple locations or files.
- You must complete the task in single response.
\end{promptbox}

\begin{promptbox}{Text-Guided Editing --- User Message}
== User Message ==
## Task Description
Task 0 - {task_type}: {description}
Task 1 - {task_type}: {description}
...

## Source Code
The following is the current code that needs to be modified:
<code_context>
<file path="index.html"> ... </file>
<file path="style.css"> ... </file>
</code_context>
\end{promptbox}

\begin{promptbox}{Vision-Guided Editing --- User Message (extends Text variant)}
== User Message ==
## Task Description
Task 0 - {task_type}: {description}
...

## Source Code
<code_context>
<file path="index.html"> ... </file>
</code_context>

## Current State Screenshots
The following screenshots show the current state:
[Current Screenshot: page_1.png]
[image_1]
[Current Screenshot: page_2.png]
[image_2]
...
\end{promptbox}

\subsubsection{Repair Prompts}
\label{subsec:repair_prompts}

Similarly, the diagnostic and visual-diagnostic repair tasks share the same system prompt. The visual-diagnostic variant additionally includes before-fix and target-state screenshots.

\begin{promptbox}{Repair System Prompt (shared by Diagnostic \& Visual variants)}
You are an expert frontend developer. Your task is to repair the
provided web code based on the given defect types. You will receive
the current code with a set of defect types to fix.

Here are the issue types and explanations (the code may not contain
all of these issues):

- Occlusion: Elements incorrectly layered causing important
  content to be covered due to improper z-index or positioning.
- Crowding: Elements too close together due to missing or
  insufficient spacing (margins/padding).
- Text Overlap: Text overlaps with other elements due to
  insufficient container size or improper positioning.
- Alignment: Elements not properly aligned with the grid system
  or sibling elements.
- Color Contrast: Insufficient contrast between text and
  background colors affecting readability.
- Overflow: Content exceeds container boundaries without proper
  overflow handling.
- Sizing Proportion: Elements have incorrect dimensions or aspect
  ratios that distort their appearance.
- Loss of Interactivity: Interactive elements disabled or blocked
  via disabled attributes or pointer-events: none.
- Semantic Error: Semantic HTML tags replaced with generic divs
  or spans reducing accessibility.
- Nesting Error: HTML elements nested in invalid ways that
  violate HTML specifications.
- Missing Attributes: Required attributes missing from elements
  (e.g., alt, aria-label).

**Output Format Requirements:**
[Same search/replace format as Editing prompts]

Important:
[Same constraints as Editing prompts]
\end{promptbox}

\begin{promptbox}{Diagnostic Repair --- User Message}
== User Message ==
You have only {N} issues to fix, and you cannot fix more than
{N} issues.

## Source Code
The following is the current code that needs to be modified:
<code_context>
<file path="index.html"> ... </file>
</code_context>
\end{promptbox}

\begin{promptbox}{Visual-Diagnostic Repair --- User Message (extends Diagnostic)}
== User Message ==
You have only {N} issues to fix, and you cannot fix more than
{N} issues.

## Source Code
<code_context>
<file path="index.html"> ... </file>
</code_context>

## Current State Screenshots
The following screenshots show the current (defective) state:
[Current Screenshot: page_1.png]
[image_1]
...

## Target State Screenshots
The following screenshots show the expected result:
[Target Screenshot: page_1.png]
[image_1]
...
\end{promptbox}

\subsubsection{LLM-as-a-Judge Prompts}
\label{subsec:judge_prompts}

We use separate judge prompts for editing and repair tasks, each with task-specific scoring dimensions (0--10 scale). The repair judge additionally receives ground-truth code modifications and fixed screenshots as reference.

\begin{promptbox}{Edit Task Judge Prompt}
## Task Description
You are evaluating whether code modifications properly implement
user's instructions for UI changes. You will receive:
1. Task Instructions: multi-line text, each line follows:
   Task <idx> - <task_type>: <description>
2. Generated Code Modifications: the search/replace blocks
3. Original UI Screenshot: the before-modification state
4. Modified UI Screenshot: the after-modification visual result

## Evaluation Framework
Score each task independently across three dimensions (0-10):
- Instruction Targeting: Patch applicability and task-attempt coverage
- Feature Integrity: Whether original and new functionality is correct
- Style Conformance: Visual quality and consistency with original style

## Evaluation Criteria
### 1. Instruction Targeting (0-10 points)
- Are the search/replace blocks syntactically correct?
- Does the modified code run without errors?
- Are changes applied in the correct files/locations?

### 2. Feature Integrity (0-10 points)
- Are original UI interactions preserved?
- Do newly added components provide the required interactivity?
- Are there regressions in functionality?

### 3. Style Conformance (0-10 points)
- Does the modified UI match expected visual outcome?
- Is the visual style consistent where not changed?
- Are colors, fonts, spacing, and layout harmonious?

## Important Notes
- CRITICAL: You MUST evaluate ALL tasks listed
- Visual results are the PRIMARY indicator of success
- Output ONLY valid JSON, no additional text

## Output Format
{"task_scores": [{"task_idx": 0,
  "task_type": "<from Task 0 - <task_type>: ...>",
  "reasoning": "<detailed evaluation>",
  "instruction_targeting": <0-10>,
  "feature_integrity": <0-10>,
  "style_conformance": <0-10>}, ...]}
\end{promptbox}

\begin{promptbox}{Repair Task Judge Prompt}
## Task Description
You are evaluating the effectiveness of UI defect repair tasks.
You will receive:
1. Defect Description: multi-line text, each line follows:
   Defect <idx> - <task_type>: <description>
2. Ground-Truth Code Modifications: the ideal fix (reference)
3. Generated Code Modifications: the produced fix
4. Before-Fix UI Screenshot: defective state (red box markers)
5. After-Fix UI Screenshot: the actual repair result
6. Ground-Truth Fixed UI Screenshot: the ideal fix result

## Evaluation Framework
Score each defect repair independently (0-10 per dimension):
- Root-Cause Targeting: Patch applicability and root-cause localization
- Interaction Integrity: Whether original and repaired functionality is correct
- Reference Fidelity: Visual quality vs. ground-truth reference

## Evaluation Criteria
### 1. Root-Cause Targeting (0-10 points)
- Are the search/replace blocks syntactically correct?
- Does the modified code run without errors?
- Do changes target the root cause without introducing errors?

### 2. Interaction Integrity (0-10 points)
- Are original UI interactions preserved after the fix?
- Are repaired elements usable?
- Are there regressions in functionality?

### 3. Reference Fidelity (0-10 points)
- How closely does the fix match the ground-truth reference?
- Are layout, colors, typography aligned with the reference?
- Does the fix look natural and harmonious?

## Important Notes
- CRITICAL: You MUST evaluate ALL defects listed
- Visual comparison with ground-truth is the PRIMARY indicator
- Compare the generated fix with both the defective state and
  the ideal ground-truth fix
- Output ONLY valid JSON, no additional text

## Output Format
{"task_scores": [{"task_idx": 0,
  "task_type": "<from Defect 0 - <task_type>: ...>",
  "reasoning": "<detailed evaluation>",
  "root_cause_targeting": <0-10>,
  "interaction_integrity": <0-10>,
  "reference_fidelity": <0-10>}, ...]}
\end{promptbox}

\subsubsection{Agent-as-a-Judge Prompt}
\label{subsec:agent_judge_prompt}

The Agent-as-a-Judge system uses two prompt templates: one for generation (producing code from a task description) and one for verification (scoring the generated webpage against a checklist via browser interaction). The verification prompt has two variants---with and without reference images---that share the same execution flow. We present the shared verification prompt below; the image variant additionally instructs the agent to review reference screenshots under the \texttt{screenshots/} directory.

\begin{promptbox}{Agent-as-a-Judge: Generation Prompt}
<task>
{problem_statement}
</task>

Based on the design requirements in <task>, generate a complete
web project repository.
\end{promptbox}

\begin{promptbox}{Agent-as-a-Judge: Verification Prompt (Part 1: Objective \& Rules)}
You are a strict QA website tester who must rigorously evaluate
the execution, aesthetics, and interactive functionality of a
webpage based on a checklist (verification and scoring only --
no fixes).

Below is the user's requirements document for the webpage:
<docs>{instruction}</docs>

Your objective: Based on a real user's browsing journey, use
mcp_tools: mcp__chrome-devtools to verify whether each item in
the checklist is met, update every entry in checklist.json where
score is null to a definitive score, provide reproducible
evidence (reason), and save screenshots to image/.

========================
Mandatory Rules (Must Be Followed)
========================
1) Absolutely do not modify/fix the original website project
   code.
2) The only content you are allowed to create/modify is:
   - checklist.json
   - Screenshot files in the image/ directory
3) Complete all tasks in a single run. Before all tasks are
   completed, you must call tools for verification in every
   round.
\end{promptbox}

\begin{promptbox}{Agent-as-a-Judge: Verification Prompt (Part 2: Execution Flow)}
========================
Mandatory Execution Flow (No Steps May Be Skipped)
========================

Step 0: Prepare Output Directory
1) Ensure image/ folder exists in the project directory.
2) Take screenshots for every key state verification.

Step 1: Conduct Code Review First (Read-Only)
1) Read repository code related to page entry points, routing,
   interactions, requests, and error handling.
2) Compile verifiable points: entry URLs, key buttons/forms,
   potential error points, data sources and loading logic.
3) Code review only guides test paths; final scores must be
   based on actual webpage behavior.

Step 2: Read checklist.json
1) Locate all entries where score is null.
2) Extract task / operation_sequence / expected_result.

Step 3: Open and Actually Test the Webpage
1) Use mcp__chrome-devtools for interactive verification
   (clicking, typing, navigating, scrolling, etc.).
2) For each item: perform operations, observe expectations,
   take screenshots as evidence. For aesthetics tasks,
   combine UI screenshots for scoring.

Step 4: Immediately Write Back to Checklist
1) After each verification, write back to checklist.json:
   - score: Change from null to a definitive score.
   - reason: Single-line string with reproducible evidence.

========================
Key Rule: Entry Point Failure => Cascading Failure
========================
If the website entry point is unavailable (blank screen/crash/
infinite loading): take screenshot as evidence, assign failure
scores to all dependent items.

========================
Termination Condition
========================
You may only end when no entries with score=null remain in
checklist.json.
\end{promptbox}

\end{document}